\documentclass[12pt,fleqn]{article}
\usepackage{epsf}
\makeatletter
\@addtoreset{equation}{section}
\renewcommand\theequation{\thesection.\arabic{equation}}
\renewcommand\appendix{\par
  \setcounter{section}{0}
  \renewcommand\thesection{Appendix \Alph{section}}
  \renewcommand\theequation{\Alph{section}.\arabic{equation}}}
\renewenvironment{titlepage}{
  \setcounter{page}{0}
  \thispagestyle{empty}}
  {\newpage}
\long\def\@makecaption#1{%
  \vskip\abovecaptionskip
  \sbox\@tempboxa{#1}%
  \ifdim \wd\@tempboxa >\hsize
    #1\par
  \else
    \hbox to\hsize{\hfil\box\@tempboxa\hfil}%
  \fi
  \vskip\belowcaptionskip}
\makeatother
\begin{document}
\newcommand{\pub}[2]{
  \hfill INLO--PUB--#1/#2
  \vspace*{1.5cm}}
\newcommand{\institute}{
  Instituut--Lorentz\\
  University of Leiden\\
  P.O. Box 9506\\
  2300 RA Leiden\\
  The Netherlands}
\renewcommand{\title}[3]{
  \begin{center}
    \large{\bf #1}\\[1.8cm]
    #2\\[1.8cm]
    \institute\\[1.5cm]
    #3\vspace*{2cm}
  \end{center}}
\renewcommand{\abstract}[1]{\noindent {\bf Abstract}\\[8mm] \noindent #1}
\begin{titlepage}
\pub{14}{94}
\title{Order $\alpha_s^{2}$ contributions to the Drell-Yan cross section
       at fixed target energies}
      {P.J. Rijken and W.L. van Neerven}
      {August 1994}
\abstract{
{}From the literature one infers that the bulk of the order $\alpha_{s}$
corrections to the Drell-Yan cross sections $d\sigma/dm$ and
$m^{3}d^{2}\sigma/dmdx_{F}$ is constituted by the soft plus virtual
gluon part of the coefficient function. In the case of $d\sigma/dm$
it can be shown that at fixed target energies the effect of the exact order
$\alpha_{s}^{2}$ corrected coefficient function is very well approximated
by its soft plus virtual gluon part. Since the complete order $\alpha_{s}^{2}$
contribution to the coefficient function is missing we have to assume
that the
same approximation also holds for $m^{3}d^{2}\sigma/dmdx_{F}$.
It appears that the
discrepancy between the exact order $\alpha_{s}$ corrected cross section
and the massive lepton pair data taken at fixed target experiments
can be partially explained by
including the order $\alpha_{s}^{2}$ soft plus virtual gluon part
of the coefficient function.}
\end{titlepage}
\section{Introduction}
Massive lepton pair production in hadronic interactions is besides
deep inelastic lepton-hadron scattering one of the most important
probes of the structure of hadrons. It is well established that one
of the dominant production mechanisms is the Drell-Yan (DY) process
\cite{r1}. Here the lepton pair is the decay product of one of the
electroweak vector bosons of the standard model ($\gamma^{\ast},
W$ and $Z$) which in the Born approximation are produced by the
annihilation of quarks and anti-quarks coming from the colliding hadrons.
This process is of experimental interest because it provides us with
an alternative way to measure the parton densities of the proton and
neutron which have been very accurately determined by the deep inelastic
lepton hadron experiments. Moreover it enables us to measure the parton
densities of unstable hadrons like pions and kaons which is impossible
in deep inelastic lepton-hadron scattering. Besides the measurement
of the parton densities there are other important tests of perturbative
quantum chromo dynamics (QCD) which can be carried out by studying the DY
process.
Here we want to mention the scale evolution of the parton densities,
although not observed in this process because of the low statistics,
and the measurement of the running coupling constant $\alpha_s(\mu^{2})$
which includes the QCD scale $\Lambda$. Finally this process constitutes
an important background for other production mechanisms of lepton pairs.
Examples are $J/\Psi$ and $\Upsilon$ decays or thermal emission
of lepton pairs in heavy-ion collisions \cite{r2}.\\
The DY process is also of theoretical interest. Since it is one
of the few reactions which can be calculated up to second order in
perturbation theory it enables us to study the origin of
large QCD corrections which
are mostly due to soft gluon bremsstrahlung and virtual gluon
contributions.
In order to control these corrections in the perturbation series one
has constructed various kinds of resummation techniques mostly leading
to the exponentiation of the dominant terms \cite{r3}-\cite{r7}.
Another issue is the dependence of the physical quantities on the chosen
scheme and the choice of scales. Since the perturbation series is truncated
the theoretical cross section will depend on the scheme and the
renormalization/factorization scale $\mu$. These dependences can be
reduced by including higher order terms in the perturbation series.
An alternative way is to determine $\mu$ itself (optimum scale) by
using so called improved perturbation theory like the principle of
minimal sensitivity (PMS) \cite{r8}, fastest apparent convergence (FAC)
\cite{r9} or the Brodsky-Lepage-Mackenzie (BLM) procedure \cite{r10}.\\
The first fixed target experiment on massive lepton pair production was
carried out by the Columbia-BNL group \cite{r11}.
Later on this process was studied in many other experiments which
were carried out at increasing energies (for reviews see \cite{r12}).
When the statistics of the data was improving one discovered that the
cross section could not be described by the simple parton model given
by S.D. Drell and T.M. Yan in \cite{r1}. This was revealed for the first
time by the NA3 experiment \cite{r13} (see also \cite{r13_2}) where the data
show a discrepancy
in the normalization between the experimental and theoretical cross section.
This discrepancy is
expressed by a so called $K$-factor which is defined by the ratio between
the experimentally observed cross section and its theoretical prediction.
The above group and the experiments carried out later on \cite{r14} show
that this $K$-factor ranges between 1.5 and 2.5 and is roughly independent
of the type of incoming hadrons. The most generally accepted explanation
of this $K$-factor was provided by perturbative QCD. The calculation
of the order $\alpha_{s}$ corrections \cite{r15}-\cite{r18} to the DY
cross section in \cite{r1} show that a considerable part of the $K$-factor
can be attributed to next-to-leading order effects. However the order
$\alpha_{s}$ corrections do not account for the whole $K$-factor. More
recent experiments \cite{r19}-\cite{r22} still indicate that the ratio
between the experimental cross section and the order $\alpha_{s}$ corrected
theoretical prediction is about 1.4, a number which might be explained by
including QCD corrections beyond order $\alpha_{s}$ as we
will show in this paper.\\
As has been mentioned at the beginning the DY process is one of the
few processes where the order $\alpha_{s}^{2}$ corrections to the
coefficient function are completely known. The latter refers to the
cross section $d\sigma/dm$ only where $m$ denotes the
lepton pair invariant mass. This coefficient function has been calculated
in the $\overline{\mbox{MS}}$ \cite{r23} as well as in the DIS \cite{r24}
scheme. However in the case of the double differential cross section
$d^{2}\sigma/dmdx_{F}$ ($d^{2}\sigma/dmdy$) one has only calculated the
order $\alpha_{s}^{2}$ part of  the coefficient function
which is due to soft and virtual gluon contributions \cite{r25} because the
remaining part is very complicate to compute. Fortunately as is shown in the
literature \cite{r15}-\cite{r18} the soft plus virtual gluon corrections
dominate the total and differential
DY cross sections in particular at fixed target energies so that we can
restrict to them to make reliable predictions.\\
An analysis of the higher order corrections to the total DY cross section
for $W$- and $Z$-production at large hadron collider energies
has been performed
in \cite{r23,r24}. Such an analysis is still missing for the DY process at
fixed target energies and therefore we present it here. In particular we want
to show that the discrepancy in the normalization between the order
$\alpha_{s}$ corrected DY cross section and the one measured at the
fixed target experiments can be partially explained by including
the order $\alpha_{s}^{2}$ contributions due to soft plus virtual
gluon effects.\\
This paper is organized as follows. In section 2 we present the
expressions for the various DY cross sections and give a review of
the partonic subprocesses included in our analysis. In section 3 the
validity of the soft plus virtual gluon approximation will be discussed
and we make a comparison between the order $\alpha_{s}^{2}$ corrected
cross section and the most recent fixed target DY data. In appendices
A and B we give the coefficient functions for $d^{2}\sigma/dmdx_{F}$
($d^{2}\sigma/dmdy$) corrected up to order $\alpha_{s}$ and order
$\alpha_{s}^{2}$ respectively. They are presented for arbitrary
renormalization and mass factorization scale in the $\overline{
\mbox{MS}}$- as well as in the DIS-scheme.
\newpage
\section{Higher order QCD corrections to $d^{2}\sigma/dmdx_{F}$
         \newline ($d^{2}\sigma/dmdy$) and $d\sigma/dm$}
Massive lepton pair production in hadron-hadron collisions proceeds
through the following reaction
\begin{eqnarray}
  H_{1}+H_{2} \;\rightarrow & V & +\hspace{3mm} "X" \nonumber \\
  & \hspace{5pt}\lfloor\raisebox{-5.9pt}{$\hspace{-2pt}\rightarrow$}
  & \raisebox{-5.5pt}{$\ell_{1} + \ell_{2}$}
\end{eqnarray}
Here $H_{1}$ and $H_{2}$ denote the incoming hadrons and $V$ is one
of the vector bosons of the standard model ($\gamma^{\ast}$,$Z$ or $W$) which
subsequently decays into a lepton pair ($\ell_{1}$,$\ell_{2}$). The symbol
$"X"$ denotes any inclusive hadronic final state which is allowed by
conservation of quantum numbers. Following the QCD improved parton model
as originally developed in \cite{r1} the double differential DY cross section
can be written as
\begin{eqnarray}
  \frac{d^{2}\sigma}{dQ^{2}dx_{F}} = \sum_{i,j} \sigma_{V}(Q^{2},M_{V}^{2})
   \,\int_{x_{1}}^{1}dt_{1}\,\int_{x_{2}}^{1}dt_{2}\,
   H_{ij}(t_{1},t_{2},\mu^{2})
   \Delta_{ij}(t_{1},t_{2},x_{1},x_{2},Q^{2},\mu^{2}).\hspace*{-2cm}
   && \nonumber \\
  &&
  \label{diff_cross}
\end{eqnarray}
Here $Q^{2}=m^{2}$ where $m$ denotes the lepton pair invariant mass.
The longitudinal momentum fraction $x_{F}$ of the lepton pair and
the Bj{\o}rken scaling variable are defined by
\begin{equation}
  x_{F}=x_{1}-x_{2}=\frac{2p_{L}}{\sqrt{S}},\hspace{2cm}
  \tau=\frac{Q^{2}}{S}=x_{1}x_{2},
  \label{def_xf_tau}
\end{equation}
where $\sqrt{S}$ stands for the center of mass energy of the incoming
hadrons $H_{1}$ and $H_{2}$. The quantity $\sigma_{V}$ is the pointlike
DY cross section which describes the process
\begin{equation}
  q_{1}+\bar{q}_{2} \rightarrow V \rightarrow \ell_{1}+\ell_{2},
\end{equation}
where $q_{1}$ and $\bar{q}_{2}$ denote the incoming quark and
anti-quark respectively. If we limit ourselves to $V=\gamma^{\ast}$,$Z$ then
$\sigma_{V}$ gets the form
\begin{eqnarray}
  \sigma_{V}(Q^{2},M_{Z}^{2}) &=& \frac{4\pi\alpha^{2}}{9Q^{4}}
    \left[ e_{\ell}^{2}e_{q}^{2} +
    \frac{2Q^{2}(Q^{2}-M_{Z}^{2})}{|Z(Q^{2})|^{2}}e_{\ell}e_{q}
    C_{V,\ell}C_{V,q} \right. \nonumber \\[2ex]
  && \left.+\frac{(Q^{2})^{2}}{|Z(Q^{2})|^{2}}(C_{V,\ell}^{2}
    +C_{A,\ell}^{2})(C_{V,q}^{2}+C_{A,q}^{2})\right]
  \label{sigma_V}
\end{eqnarray}
with
\begin{equation}
  Z(Q^{2}) = Q^{2}-M_{Z}^{2}+i M_{Z} \Gamma_{Z}.
\end{equation}
Here the width of the $Z$-boson is taken to be energy independent and
all fermion masses are neglected since they are much smaller than
$\sqrt{Q^{2}}$. The charges of the leptons and quarks are given by
\begin{equation}
  e_{\ell}=-1,\hspace{2cm}e_{u}=\frac{2}{3},
  \hspace{2cm}e_{d}=-\frac{1}{3}.
\end{equation}
The vector- and axial-vector coupling constants of the $Z$-boson to the
leptons and quarks are equal to
\begin{equation}
  \begin{array}{ll}
    \displaystyle
    C_{A,\ell} = \frac{1}{2\sin 2\theta_{W}} &
    C_{V,\ell} = -C_{A,\ell}(1-4\sin^{2}\theta_{W}) \\[2ex]
    C_{A,u} = -C_{A,d} = -C_{A,\ell} & \\[2ex]
    C_{V,u} = C_{A,\ell} \left(1-\frac{8}{3}\sin^{2}
     \theta_{W}\right) &
    C_{V,d} = -C_{A,\ell} \left(1-\frac{4}{3}\sin^{2}
     \theta_{W}\right).
  \end{array}
\end{equation}
The function $H_{ij}$ in (\ref{diff_cross}) stands for the combination of
parton densities corresponding to the incoming partons $i$ and $j$ ($i,j=
q,\bar{q},g$). Finally $\Delta_{ij}$ denotes the DY coefficient
function which is determined by the partonic subprocess
\begin{equation}
  i + j \rightarrow V + "X"
\end{equation}
where $"X"$ now represents any multi partonic final state. Both functions
$H_{ij}$ and $\Delta_{ij}$ depend in addition to the scaling
variables $t_{i}$ and $x_{i}$ also on the renormalization and mass
factorization scales which are usually put to be equal to $\mu$.
Besides the cross section in (\ref{diff_cross}) one is sometimes also
interested in the rapidity distribution of the lepton pair. In this case
the left hand side in (\ref{diff_cross}) is replaced by $d^{2}\sigma/dQ^{2}dy$
where $y$ denotes the rapidity defined by (see (\ref{def_xf_tau}))
\begin{equation}
  y = \frac{1}{2}\ln\frac{x_{1}}{x_{2}},\hspace{1cm}
  x_{1} = \sqrt{\tau}\,e^{\displaystyle y},\hspace{1cm}
  x_{2} = \sqrt{\tau}\,e^{\displaystyle -y}
\end{equation}
or
\begin{equation}
  y = \frac{1}{2}\ln\frac{ x_{F}+\sqrt{x_{F}^{2}+4\tau}}
                         {-x_{F}+\sqrt{x_{F}^{2}+4\tau}}.
\end{equation}
Furthermore on the right hand side the coefficient function $\Delta_{ij}$
is replaced by its analogue corresponding to the cross section
$d^{2}\sigma/dQ^{2}dy$.\\
The coefficient function $\Delta_{ij}$ (\ref{diff_cross}) can be expanded
as a power
series in the running coupling constant $\alpha_{s}(\mu^{2})$ as follows
\begin{equation}
  \Delta_{ij} = \sum_{n=0}^{\infty}\left(\frac{\alpha_{s}(\mu^{2})}
    {4\pi}\right)^{n}\,\Delta_{ij}^{(n)}.
  \label{def_deltas}
\end{equation}
In lowest order the coefficient function of the differential cross
section (\ref{diff_cross}) is determined by the subprocess
\begin{equation}
  q + \bar{q} \rightarrow V.
  \label{born_qqv}
\end{equation}
Here $V$ either stands for the virtual photon $\gamma^{\ast}$ or the
$Z$-boson and the coefficient function is given by
\begin{equation}
  \Delta_{q\bar{q}}^{(0)} = \frac{1}{x_{1}+x_{2}}
    \delta(t_{1}-x_{1})\,\delta(t_{2}-x_{2}).
  \label{delta0_qqb}
\end{equation}
The order $\alpha_{s}$ corrections to the Born process (\ref{born_qqv})
denoted by $\Delta_{q\bar{q}}^{(1)}$  are given by the one-loop
contributions to (\ref{born_qqv}) and the gluon bremsstrahlung process
\begin{equation}
  q + \bar{q} \rightarrow V + g.
  \label{qqvg}
\end{equation}
In addition to the process above we have another reaction which instead
of a quark or anti-quark has a gluon in the initial state
\begin{equation}
  g + q(\bar{q}) \rightarrow V + q(\bar{q}).
  \label{gqvq}
\end{equation}
This reaction contributes to $\Delta_{gq}^{(1)}$. Both contributions
$\Delta_{q\bar{q}}^{(1)}$ and $\Delta_{gq}^{(1)}$ have been calculated
in \cite{r16,r17,r26} (DIS-scheme) and
in \cite{r27} ($\overline{\mbox{MS}}$-scheme) and are presented in
(\ref{d_qqb_1}) and (\ref{d_gq_1}), (\ref{d_qg_1}) respectively.
A part of the order $\alpha_{s}^{2}$ corrections to the coefficient
function corresponding to $d^{2}\sigma/dQ^{2}dx_{F}$ has also been
calculated in~\cite{r25}. These corrections originate from the
soft plus virtual gluon contributions. They consist of the two-loop
corrections to process (\ref{born_qqv}) and the one-loop correction to
process (\ref{qqvg}) where the gluon is taken to be soft. Furthermore
one has also included the bremsstrahlungs process
\begin{equation}
  q + \bar{q} \rightarrow V + g + g
  \label{qqvgg}
\end{equation}
and fermion pair production
\begin{equation}
  q + \bar{q} \rightarrow V + q + \bar{q}
  \label{born_qqbvqqb}
\end{equation}
where the gluons were taken to be soft and the quark--anti-quark pair
in the final state of (\ref{born_qqbvqqb}) has a low invariant mass.\\
All above corrections contribute to $\Delta_{q\bar{q}}^{(2)}$ and can be
found in appendix B for arbitrary factorization and renormalization
scale $\mu$ where they are presented in the $\overline{\mbox{MS}}$- as well
as in the DIS-scheme.
The hard gluon corrections (\ref{qqvgg}) and the other two-to-three
body processes (see below) are very hard to compute at least for the
double differential cross sections.
Fortunately as has been shown in \cite{r15}-\cite{r18} the bulk of
the order $\alpha_{s}$ radiative corrections to the cross sections
$d\sigma/dQ^{2}$ and $d^{2}\sigma/dQ^{2}dx_{F}$ is constituted by the soft plus
virtual gluon contributions to $\Delta_{q\bar{q}}^{(1)}$.
Therefore within the experimental and theoretical uncertainties one can
assume that the order $\alpha_{s}^{2}$ part of the coefficient function
$\Delta_{q\bar{q}}$ which is only due to soft plus virtual gluon
contributions is
sufficient to describe the next-to-next-to-leading order DY cross section at
fixed target energies. This can be tested for the quantity $d\sigma/dQ^{2}$
which is defined by
\begin{equation}
\begin{tabular}{ccc}
  $ \displaystyle
    \frac{d\sigma}{dQ^{2}} = \int_{\tau-1}^{1-\tau}dx_{F}\,
    \frac{d^{2}\sigma}{dQ^{2}dx_{F}} $
  & or &
  $ \displaystyle
    \frac{d\sigma}{dQ^{2}} = \int_{\frac{1}{2}\ln\tau}^{-\frac{1}{2}
    \ln\tau}dy\,\frac{d^{2}\sigma}{dQ^{2}dy} $,
\end{tabular}
\label{dsdq}
\end{equation}
which can also be written as
\begin{equation}
  \frac{d\sigma}{dQ^{2}} = \sum_{i,j} \sigma_{V}(Q^{2},M_{V}^{2})
  \int_{\tau}^{1}\frac{dx_{1}}{x_{1}}\int_{\frac{\tau}{x_{1}}}^{1}
  \frac{dx_{2}}{x_{2}}\,H_{ij}(x_{1},x_{2},\mu^{2})
  \Delta_{ij}\left(\frac{\tau}{x_{1}x_{2}},\frac{Q^{2}}{\mu^{2}}\right)
  \label{dsdq_2}
\end{equation}
where $\Delta_{ij}$ now stands for the coefficient function corresponding
to the integrated cross section $d\sigma/dQ^{2}$.\\
Since the exact order $\alpha_{s}^{2}$ corrections to this coefficient
function are completely known see \cite{r23} ($\overline{\mbox{MS}}$-scheme)
and \cite{r24} (DIS-scheme) one can now make a comparison between the exact DY
cross section coming from the complete coefficient function and the
approximate cross section due to the soft plus virtual gluon part. The full
order $\alpha_{s}^{2}$ contribution to the DY coefficient function requires
besides the calculation of the subprocesses mentioned above the computation of
the following two-to-three body partonic subprocesses. First we have the
bremsstrahlungs correction to (\ref{gqvq})
\begin{equation}
  g + q(\bar{q}) \rightarrow V + q(\bar{q}) + g
  \label{gqvqg}
\end{equation}
which entails the computation of the one-loop corrections to
(\ref{gqvq}). In addition one has to add the subprocesses
\begin{equation}
  q_{1} + \bar{q}_{2} \rightarrow V + q_{1} + \bar{q}_{2}
  \label{born_q1q2vq1q2} \\[2ex]
\end{equation}
\begin{equation}
  q(\bar{q}) + q(\bar{q}) \rightarrow V + q(\bar{q}) + q(\bar{q})
  \label{born_qqvqq}
\end{equation}
and
\begin{equation}
  g + g \rightarrow V + q + \bar{q}.
  \label{born_ggvqq}
\end{equation}
Reactions (\ref{gqvqg}),(\ref{born_q1q2vq1q2}),(\ref{born_qqvqq}) and
(\ref{born_ggvqq}) contribute to the coefficient functions
$\Delta_{gq}^{(2)}$, $\Delta_{q\bar{q}}^{(2)}$,
$\Delta_{qq}^{(2)}$ and $\Delta_{gg}^{(2)}$ respectively. The exact
result of the coefficient function calculated up to order $\alpha_{s}^{2}$
for $d\sigma/dQ^{2}$ gives an indication about the validity of the soft
plus virtual gluon approximation of $d^{2}\sigma/dQ^{2}dx_{F}$ (or
$d^{2}\sigma/dQ^{2}dy$) for which a complete order $\alpha_{s}^{2}$
calculation is still missing. In \cite{r24} one has made a detailed analysis
of this approximation for the total cross section of $W$- and $Z$-
production which is derived from (\ref{dsdq_2}) by integrating $d\sigma/dQ^{2}$
over $Q^{2}$. From this analysis one infers that the approximation works
quite well in
order $\alpha_{s}$ as well as in order $\alpha_{s}^{2}$ when $M_{V}^{2}/S >
0.01$ provided the DY coefficient function is computed in the DIS-scheme.
This implies
that in practice one can only apply it to the cross section measured at the
$\mbox{Sp}\bar{\mbox{p}}\mbox{S}$ ($\sqrt{S} = 0.63\,\mbox{TeV}$). The reason
that this happens in the DIS-scheme is purely accidental. It originates
from the large coefficient of the delta-function $\delta(1-x)$ appearing in
$\Delta_{q\bar{q}}(x)$ which is small in the $\overline{\mbox{MS}}$-scheme.
Apparently the combination of the anomalous dimension (Altarelli-Parisi
splitting function) and the remaining part of the coefficient function is
very small in the DIS-scheme. It is expected that the approximation will
even work better when $\tau=Q^{2}/S \rightarrow 1$, a condition which is
satisfied by fixed target experiments. In this case the phase space of the
multi partonic final state in the above reactions will be reduced so that only
soft gluons or fermion pairs with low invariant mass can be radiated off.
Their contributions manifest themselves by
large logarithms of the type $(\ln^{k}(1-x)/(1-x))_{+}$ which appear in
the coefficient function in the DIS- as well as in the
$\overline{\mbox{MS}}$-scheme.\\
Notice that the above analysis holds if the mass factorization scale $\mu$ is
chosen to be $\mu^{2}=Q^{2}$. Therefore it is not impossible that the above
conclusions have to be altered when a scale completely different from
$\mu^{2}=Q^{2}$ is adopted.\\
Finally one has to bear in mind that a complete next-to-next-to-leading order
analysis cannot be carried out yet because the appropriate parton densities
are not available. The latter can be attributed to the fact that the
three-loop contributions to the Altarelli-Parisi splitting functions or the
anomalous dimensions have not been calculated up to so far. Therefore the
analysis of the order $\alpha_{s}^{2}$ corrected result for $d\sigma/dQ^{2}$
has to be considered with caution. This holds even more for the order
$\alpha_{s}^{2}$ corrected differential distribution
$d^{2}\sigma/dQ^{2}dx_{F}$ or $d^{2}\sigma/dQ^{2}dy$.
\newpage
\section{Results}
In this section we start with a discussion of the validity of the soft plus
virtual gluon ($S+V$) approximation of the order $\alpha_{s}^{2}$
correction to $d^{2}\sigma/dQ^{2}dx_{F}$
(\ref{diff_cross}). This is done by making a comparison with the integrated
cross
section $d\sigma/dQ^{2}$ (\ref{dsdq}) for which the coefficient function is
completely known up to order $\alpha_{s}^{2}$. Then we include this
approximation in our analysis of the fixed target
muon pair data published in \cite{r19}-\cite{r22}. In particular we show
that this correction partially accounts for the difference in the
normalization between the data in \cite{r19}-\cite{r22} and the order
$\alpha_{s}$
corrected cross section calculated in \cite{r16,r17,r26,r27}.\\
The calculation of the cross sections $d\sigma/dQ^{2}$ (\ref{dsdq}) and
$d^{2}\sigma/dQ^{2}dx_{F}$ (\ref{diff_cross}) will be performed in the
DIS- as well as in the
$\overline{\mbox{MS}}$-scheme chosen for the coefficient functions as well as
for the parton densities. The coefficient functions for $d\sigma/dQ^{2}$ up to
order $\alpha_{s}^{2}$ can be found in \cite{r23}
($\overline{\mbox{MS}}$-scheme) and \cite{r24} (DIS-scheme). The coefficient
functions for $d^{2}\sigma/dQ^{2}dx_{F}$ corrected up to order $\alpha_{s}$ are
obtained from \cite{r16,r17,r26} (DIS-scheme) and \cite{r27}
($\overline{\mbox{MS}}$-scheme). In order to make this paper self-contained we
have also presented them in appendix A. The order
$\alpha_{s}^{2}$ contribution as far as the soft plus virtual gluon
part is concerned has been calculated in \cite{r25} and is
presented in both schemes in a more amenable form in appendix B. For the
next-to-leading order nucleon parton densities we have chosen the MRS(D-) set
\cite{r28} for which a DIS- ($\Lambda = 230\,\mbox{MeV}$) and an
$\overline{\mbox{MS}}$-version ($\Lambda = 215\,\mbox{MeV}$) exist.
Further we use the two-loop ($\overline{\mbox{MS}}$-scheme) corrected running
coupling constant with the number of light flavors $n_{f}=4$ and the QCD
scale is the same as chosen for the MRS(D-) set. For the pion densities we take
the leading log parametrization (DO1) in \cite{r29}. Using this set one could
only fit the old lepton-pair data (for references see \cite{r29}) by allowing
an arbitrary normalization (or $K$-factor) with respect to the leading order
theoretical DY cross section. In this section it is shown that this factor
can be partially explained by including higher order QCD corrections.
Next-to-leading
(NLO) order parton densities for the pion exist in \cite{r27} and \cite{r30}
but they are
only presented in the $\overline{\mbox{MS}}$-scheme. Also here one has to
use an arbitrary $K$-factor to fit the data which is smaller than found for the
leading order process since a part of the normalization is accounted for by the
order $\alpha_{s}$ corrections. Because of the missing (NLO) parton densities
of the pion in the DIS-scheme we prefer to use the leading log parametrization
in \cite{r29}.
Finally we choose the factorization scale $\mu$ to be equal to the
renormalization scale where $\mu^{2}=Q^{2}$.
All numerical results in this paper are produced by our Fortran program DIFDY
which can be obtained on request.\\
The plots will be presented at three different fixed target energies given by
$\sqrt{S}=15.4\,;21.8\,\mbox{and}\,38.8\,\mbox{GeV}/c$. At the first energy
i.e.
$\sqrt{S}=15.4\,\mbox{GeV}/c$ one has observed muon pairs produced in the
reactions $\bar{p}+W\rightarrow\mu^{+}\mu^{-}+"X"$ and
$\pi^{-}+W\rightarrow\mu^{+}\mu^{-}+"X"$ measured by the E537 group
\cite{r19}.
The second experiment is carried out at $\sqrt{S}=21.8\,\mbox{GeV}/c$ by the
E615 \cite{r20} group where the same lepton pair is measured in the reaction
$\pi^{-}+W\rightarrow\mu^{+}\mu^{-}+"X"$. Finally we discuss the E772
experiment \cite{r21,r22} at $\sqrt{S}=38.8\,\mbox{GeV}/c$ where the reaction
$p+N\rightarrow\mu^{+}\mu^{-}+"X"$ is studied where $N$ is either
represented by the isoscalar targets $\mbox{}^{2}H$ and $C$ or by $W$
(tungsten) which has a large neutron excess. Here we will only make a
comparison with the $\mbox{}^{2}H$-data. In the case of the E537, E615
experiments $W$ is given by $Z/A = 0.405$ whereas E772 used tungsten with
$Z/A = 0.409$. Here $Z$ and $A$ denote the charge and atomic number of the
nucleus respectively. Finally notice that at the above energies we can safely
neglect the contributions coming from the Z-boson in (\ref{sigma_V}) since the
virtual photon dominates the cross section.\\
Let us first start with the discussion of the $S+V$ approximation to the
coefficient function corresponding to $d\sigma/dQ^{2}$. The soft plus
virtual gluon part of the coefficient function, which only appears in
$\Delta_{q\bar{q}}$, can be written as
\begin{eqnarray}
  \Delta_{q\bar{q}}^{S+V}(x,Q^{2},\mu^{2}) &=&
  \delta(1-x)+\sum_{i=1}^{\infty}\left(\frac{\alpha_{s}(\mu^{2})}{4\pi}
  \right)^{i}\left[\sum_{j=0}^{2i-1} a_{j}^{(i)}(Q^{2},\mu^{2})
  \left(\frac{\ln^{j}(1-x)}{1-x}\right)_{+} \right.\hspace*{-2pt}
  \nonumber \\[2ex]
  && \left.\vphantom{\frac{A}{A}} + \delta(1-x) b^{(i)}(Q^{2},\mu^{2})\right],
  \label{deltaqqb}
\end{eqnarray}
where the logarithms have to be interpreted in the distributional sense (see
\cite{r16}). The coefficients $a^{(i)}_{j}$ and $b^{(i)}$ depend on $Q^2$
and the
factorization scale $\mu^{2}$. The above coefficients can be read off the
explicit form of (\ref{deltaqqb}) given by eqs. (B.3), (B.8) in \cite{r23}
and (A.3), (A.8) in \cite{r24}. In order to test the $S+V$ approximation to
the DY cross section we study the following ratios
\begin{equation}
  R^{(1)}(\sqrt{\tau}) =
    \frac{\displaystyle
          \frac{d\sigma^{(0)}}{dm} +
          \frac{d\sigma^{S+V,(1)}}{dm}}
         {\displaystyle
          \frac{d\sigma^{(0)}}{dm} +
          \frac{d\sigma^{(1)}}{dm}}
  \label{r1vs}
\end{equation}
and
\begin{equation}
  R^{(2)}(\sqrt{\tau}) =
    \frac{\displaystyle
          \frac{d\sigma^{(0)}}{dm} +
          \frac{d\sigma^{(1)}}{dm} +
          \frac{d\sigma^{S+V,(2)}}{dm}}
         {\displaystyle
          \frac{d\sigma^{(0)}}{dm} +
          \frac{d\sigma^{(1)}}{dm} +
          \frac{d\sigma^{(2)}}{dm}}.
  \label{r2vs}
\end{equation}
In the above expressions $d\sigma^{(i)}/dm$ ($m = \sqrt{Q^{2}}$) denotes the
$O(\alpha_{s}^{i})$ contribution to the DY cross section containing the exact
$O(\alpha_{s}^{i})$ part of the coefficient function where all partonic
subprocesses are included. The quantities $d\sigma^{S+V,(i)}/dm$ stand for
the $O(\alpha_{s}^{i})$ contribution to the cross sections where only the
soft plus virtual gluon part of the coefficient function according to
(\ref{deltaqqb}) is taken into account.\\
In fig.~\ref{all_dis} we have plotted $R^{(1)}(\sqrt{\tau})$ and
$R^{(2)}(\sqrt{\tau})$ in the DIS-scheme for the $\sqrt{\tau}$-ranges
explored by the three experiments mentioned above. From the figure we infer
that the $S+V$ approximation overestimates the exact cross section by less
than $10 \%$ at small $\sqrt{\tau}$-values. At large $\sqrt{\tau}$-values
this becomes better which is to be expected since in the limit $\tau
\rightarrow 1$ the approximation becomes equal to the exact correction. In
this limit hard gluon radiation and all other partonic subprocesses like
quark-gluon scattering are suppressed because of the reduction in phase space.
By comparing $R^{(2)}(\sqrt{\tau})$ with $R^{(1)}(\sqrt{\tau})$ we observe
a slight improvement when higher order corrections are included in the
denominator as well as in the numerator. In fig.~\ref{all_ms}
we did the same as in
fig.~\ref{all_dis} but now for the $\overline{\mbox{MS}}$-scheme.
Here we observe
that the $S+V$ approximation underestimates the exact DY cross section by
more than $10 \%$ in particular when the C.M. energy $\sqrt{S}$ is small like
in the case of E537 ($\sqrt{S} = 15.4\,\mbox{GeV}/c$) or E615 ($\sqrt{S} =
21.8\,\mbox{GeV}/c$).
Furthermore $R^{(2)}(\sqrt{\tau})$ (\ref{r2vs}) becomes worse than
$R^{(1)}(\sqrt{\tau})$ (\ref{r1vs}) in particular in the low
$\sqrt{\tau}$-region. Hence we can conclude that for $d\sigma/dm$ the $S+V$
approximation works better in the DIS-scheme than in the
$\overline{\mbox{MS}}$-scheme.\\
In the case of the double differential cross section $d^{2}\sigma/dmdx_{F}$
($m = \sqrt{Q^{2}}$) the exact order $\alpha_{s}^{2}$ contribution to the
coefficient function is not known so that one can only make a comparison on
the order $\alpha_{s}$ level. The $S+V$ part of the coefficient function,
of which the explicit form is given up to order $\alpha_{s}^{2}$
in appendices A and B, becomes
\begin{eqnarray}
\lefteqn{\Delta_{q\bar{q}}^{S+V}(t_{1},t_{2},x_{1},x_{2},Q^{2},\mu^{2}) =
    \frac{1}{x_{1}+x_{2}}\left[\vphantom{\frac{A}{A}}
    \delta(t_{1}-x_{1})\,\delta(t_{2}-x_{2})\right.}
  \nonumber \\[2ex]
  &&+\sum_{i=1}^{\infty}\left(\frac{\alpha_{s}(\mu^{2})}{4\pi}\right)^{i}
   \left\{\vphantom{\frac{A}{A}}\right.\hspace*{-3mm}
   \sum_{\begin{array}{c} \scriptstyle k,l \\
           \scriptstyle k+l \leq 2i-2 \end{array}}
   a_{kl}^{(i)}(Q^{2},\mu^{2})\left(\frac{\ln^{k}(t_{1}/x_{1}-1)}
   {t_{1}-x_{1}}\right)_{+}\left(\frac{\ln^{l}(t_{2}/x_{2}-1)}
   {t_{2}-x_{2}}\right)_{+}\hspace*{-7pt}
  \nonumber \\[2ex]
  &&+ \delta(t_{1}-x_{1})\sum_{\begin{array}{c} \scriptstyle k,l \\
        \scriptstyle k+l \leq 2i-1 \end{array}}
      b_{kl}^{(i)}(Q^{2},\mu^{2})\left(\frac{\ln^{k}(t_{2}/x_{2}-1)}
      {t_{2}-x_{2}}\right)_{+}\ln^{l}\frac{1-x_{1}}{x_{1}}
  \nonumber \\[2ex]
  &&+ \delta(t_{2}-x_{2})\sum_{\begin{array}{c} \scriptstyle k,l \\
        \scriptstyle k+l \leq 2i-1 \end{array}}
      b_{kl}^{(i)}(Q^{2},\mu^{2})\left(\frac{\ln^{k}(t_{1}/x_{1}-1)}
      {t_{1}-x_{1}}\right)_{+}\ln^{l}\frac{1-x_{2}}{x_{2}}
  \nonumber \\[2ex]
  && + \delta(t_{1}-x_{1})\delta(t_{2}-x_{2})
     \sum_{\begin{array}{c} \scriptstyle k,l \\
     \scriptstyle k+l \leq 2i \end{array}}
     c_{kl}^{(i)}(Q^{2},\mu^{2})\ln^{k}\frac{1-x_{1}}{x_{1}}
     \ln^{l}\frac{1-x_{2}}{x_{2}}
     \left.\left.\vphantom{\frac{A}{A}}\right\}\right]
  \label{deltaqqbdiff}
\end{eqnarray}
where the definitions for the distributions indicated by a plus sign
can be found in appendix A.\\
To study the $S+V$ approximation we define an analogous quantity as given
for $d\sigma/dm$ in (\ref{r1vs}). In the subsequent figures we plot the ratio
\begin{equation}
  d\,R^{(1)}(\sqrt{\tau},x_{F}) =
    \frac{\displaystyle
          \frac{d^{2}\sigma^{(0)}}{dmdx_{F}} +
          \frac{d^{2}\sigma^{S+V,(1)}}{dmdx_{F}}}
         {\displaystyle
          \frac{d^{2}\sigma^{(0)}}{dmdx_{F}} +
          \frac{d^{2}\sigma^{(1)}}{dmdx_{F}}}
  \label{dr1vs}
\end{equation}
where the meaning of $d^{2}\sigma^{(i)}/dmdx_{F}$ and
$d^{2}\sigma^{S+V,(i)}/dmdx_{F}$ is the same as for $d\sigma^{(i)}/dm$ and
$d\sigma^{S+V,(i)}/dm$ defined below (\ref{r2vs}). Notice that here we cannot
present $d\,R^{(2)}(\sqrt{\tau},x_{F})$ because the exact cross section
$d^{2}\sigma^{(2)}/dmdx_{F}$ is still unknown.\\
Starting with the DIS-scheme we have plotted $d\,R^{(1)}(\sqrt{\tau},x_{F})$
at $\sqrt{S} = 15.4\,\mbox{GeV}/c$ (E537) for three representative
$\sqrt{\tau}$-values as a function of $x_{F}$ in fig.~\ref{ana_dis}. From this
figure one infers that at small $\sqrt{\tau}$ around $x_{F}=0$ the
approximate cross section overestimates the exact one by about $20\%$. This
value is much larger than in the case of the integrated cross section
$d\sigma/dm$ where it was at maximum $10\%$. The approximation becomes better
when either $|x_{F}|$ or $\sqrt{\tau}$ gets larger.\\
The overestimation is even bigger when the energy increases. This can be
observed in fig.~\ref{con_dis} ($\sqrt{S}=21.8\,\mbox{GeV}/c$, E615)
or fig.~\ref{mcg_dis}
($\sqrt{S}=38.8\,\mbox{GeV}/c$, E772). Here one overestimates the exact cross
section at small $\sqrt{\tau}$-values even by $25\%$. If we repeat our
calculations in the $\overline{\mbox{MS}}$-scheme we observe a considerable
improvement of the $S+V$ approximation to the double differential cross
section (see figs.~\ref{ana_ms}-\ref{mcg_ms}).
Although like in the case of $d\sigma/dm$
the approximation underestimates the cross section at high
$\sqrt{\tau}$-values the difference with the exact one is less than
$5\%$.\\
Summarizing our findings we conclude that in the case of the DIS-scheme the
$S+V$ approximation works better for $d\sigma/dm$ than for
$d^{2}\sigma/dmdx_{F}$ whereas for the $\overline{\mbox{MS}}$-scheme just
the opposite is happening, except for $\tau \rightarrow 1$ where
$R^{(1)}(\sqrt{\tau})$ and $d\,R^{(1)}(\sqrt{\tau},x_{F})$ become close to 1
independent of the chosen scheme. Further from
figs.~\ref{ana_dis}-\ref{mcg_ms} it appears that when
$d^{2}\sigma^{S+V,(1)}/dmdx_{F}$ is integrated over $x_{F}$ according to
(\ref{dsdq}) we get a result which differs from the one obtained from
$d\sigma^{S+V,(1)}/dm$ in (\ref{dsdq_2}) in particular at small $\sqrt{\tau}$.
On the first sight this is surprising because one expects the same cross
section $d\sigma/dm$ independent of the order of integration. However both
procedures only lead to the same answer for $d\sigma/dm$ when the full
coefficient functions are inserted in the equations for $d^{2}\sigma/dmdx_{F}$
(\ref{diff_cross}) and $d\sigma/dm$ (\ref{dsdq_2}).
If we limit ourselves to the $S+V$ part of the coefficient functions
as given in (\ref{deltaqqb}) and (\ref{deltaqqbdiff}) then the two procedures
to compute $d\sigma/dm$ only provides us with the same answer when $\tau
\rightarrow 1$. This we have also checked for the order $\alpha_{s}^{2}$
$S+V$ contribution. Therefore the expression in (\ref{deltaqqb}) is not the
integrated form of equation (\ref{deltaqqbdiff}) except if $\tau \rightarrow
1$. This explains why at large $\tau$ $R^{(1)}(\sqrt{\tau})$ (\ref{r1vs}) and
$d\,R^{(1)}(\sqrt{\tau},x_{F})$ (\ref{dr1vs}) are roughly the same and equal to
1 irrespective of the chosen scheme. The above properties of the $S+V$
approximation also reveal that if $\sqrt{\tau}$ becomes much smaller than 1
one has to be cautious in predicting the still unknown
$d\,R^{(2)}(\sqrt{\tau},x_{F})$ from the values obtained for the known
$R^{(2)}(\sqrt{\tau})$ (\ref{r2vs}) and $d\,R^{(1)}(\sqrt{\tau},x_{F})$
(\ref{dr1vs}). In the subsequent part of this work we will use as a guiding
principle that as long as $|d\,R^{(1)}(\sqrt{\tau},x_{F})-1| < 0.1$ we expect
that the $S+V$ approximation of the second order contribution to
$d^{2}\sigma/dmdx_{F}$ will be very close to the exact result. If
$|d\,R^{(1)}(\sqrt{\tau},x_{F})-1| > 0.2$ then one should not trust this
approximation and one has to rely on the predictions obtained from the first
order corrected cross section. This implies that for the experiments discussed
in this paper one can make a reasonable prediction for the second order
correction as long as $\sqrt{\tau} > 0.3$.\\
After having discussed the validity of the above approach at fixed target
energies we will now make a comparison with the data of the E537 \cite{r19},
E615 \cite{r20} and E772 \cite{r21,r22} experiments. For that purpose we
compute the Born cross section $d^{2}\sigma_{0}/dmdx_{F}$, the order
$\alpha_{s}$ corrected exact cross section $d^{2}\sigma_{1}/dmdx_{F}$ and the
order $\alpha_{s}^{2}$ corrected cross section $d^{2}\sigma_{2}/dmdx_{F}$.
Notice that in the latter only the contribution due to
the coefficient function $\Delta_{q\bar{q}}^{S+V}$
(\ref{deltaqqbdiff}) (see
appendix B) has been included because the other contributions are still
missing. The computations have been carried out in the DIS-scheme. The
results for the $\overline{\mbox{MS}}$-scheme will be shortly commented upon at
the end of this section.\\
Starting with the experiment E537 ($\sqrt{S} = 15.4\,\mbox{GeV}/c$)
we have plotted the
quantity
\begin{equation}
  \frac{d\sigma}{dm} = \int_{0}^{1-\tau}dx_{F}\,\frac{d^{2}\sigma}{dmdx_{F}}
  \label{dsdm_half}
\end{equation}
in figs.~\ref{dsdm_pbW} and~\ref{dsdm_piW} for the reactions
$\bar{p}+W\rightarrow\mu^{+}\mu^{-}+"X"$ and
$\pi^{-}+W\rightarrow\mu^{+}\mu^{-}+"X"$ respectively.
Notice that $x_{F}$ in \cite{r19} is defined
as $x_{F}=2p_{L}/[(1-\tau)\sqrt{S}]$ which differs from the usual definition
in \cite{r16,r17,r26,r27}. Since the higher order QCD corrections are
calculated for
$d^{2}\sigma/dmdx_{F}$ with $x_{F}$ defined in (\ref{def_xf_tau})
and the cross section
is not a Lorentz invariant we had to change the $x_{F}$-bins in table III of
\cite{r19} according to our definition above.
Figs.~\ref{dsdm_pbW} and~\ref{dsdm_piW} reveal
that the data are in agreement with the order $\alpha_{s}$ as well as
with the order
$\alpha_{s}^{2}$ corrected cross section but lie above the result given by
the Born approximation. The difference between the latter and the data
is observed when we consider the quantity
\begin{equation}
  \frac{d\sigma}{dx_{F}} = \int_{4.0}^{9.0}dm\,\frac{d^{2}\sigma}{dmdx_{F}}
  \label{dsdxf}
\end{equation}
which is presented in figs.~\ref{dsdxf_pbW} and~\ref{dsdxf_piW} for the above
two reactions.
Even the order $\alpha_{s}$ corrected cross section lies below the data for
$x_{F} < 0.6$ as can be seen in fig.~\ref{dsdxf_piW}. On
the other hand the order $\alpha_{s}^{2}$ corrected cross section is in
agreement with experiment over the whole $x_{F}$ range.\\
The second experiment, E615 \cite{r20} also studies the reaction
$\pi^{-}+W\rightarrow\mu^{-}\mu^{+}+"X"$ but now for $\sqrt{S}=21.8\,
\mbox{GeV}/c$. In fig.~\ref{con_dsdm} we have compared the quantity
$d\sigma/d\sqrt{\tau}=\sqrt{S}\,d\sigma/dm$
with the data where $d\sigma/dm$ is defined in the same way as in
(\ref{dsdm_half}). Apart from the bump, which is due to the $\Upsilon$
resonance at about $\sqrt{\tau}=0.43$, the order $\alpha_{s}^{2}$ corrected
cross section reasonably describes the experimental results whereas the Born
and the order $\alpha_{s}$ prediction fall below the data. The importance of
the order $\alpha_{s}^{2}$ contribution is also revealed when we study
the double differential cross section
\begin{equation}
  \frac{d^{2}\bar{\sigma}}{d\sqrt{\tau}dx_{F}} = \frac{1}{\sqrt{\tau_{2}}-
  \sqrt{\tau_{1}}}\int_{\sqrt{\tau_{1}}}^{\sqrt{\tau_{2}}}d\sqrt{\tau}
  \,\frac{d^{2}\sigma}{d\sqrt{\tau}dx_{F}}
  \label{cross_avg}
\end{equation}
for various $x_{F}$ regions, see figs.~\ref{con_bin1}-\ref{con_bin5}.
The curves predicted
by the Born and the order $\alpha_{s}$ corrections all lie below the data.
For $\sqrt{\tau} > 0.277$
even the order $\alpha_{s}^{2}$ contribution
is not sufficient to close the gap between theory and experiment.
This is due to the presence of the $\Upsilon$ in the region $0.323 <
\sqrt{\tau} < 0.599$ which has not been subtracted from the data.
The discrepancy between the order $\alpha_{s}^{2}$ corrected cross section
and the data becomes even more clear when we
plot the $K$-factor (fig.~\ref{Kfactor}) defined by
\begin{equation}
  K_{i}(\sqrt{\tau})=\frac{\displaystyle
  \int_{0}^{1-\tau}dx_{F}\,\frac{d^{2}\sigma_{i}}
  {d\sqrt{\tau}dx_{F}}}
  {\displaystyle
  \int_{0}^{1-\tau}dx_{F}\,\frac{d^{2}\sigma_{0}}{d\sqrt{\tau}dx_{F}}}
  \label{Kth}
\end{equation}
in fig.~\ref{Kfactor} and compare the above expression with the experimental
$K$-factor which is given by
\begin{equation}
  K_{\scriptstyle \mbox{exp}}(\sqrt{\tau}) = \frac{\displaystyle
  \int_{0}^{1-\tau}dx_{F}
  \,\frac{d^{2}\sigma_{\scriptstyle \mbox{exp}}}{d\sqrt{\tau}dx_{F}}}
  {\displaystyle
  \int_{0}^{1-\tau}dx_{F}\,\frac{d^{2}\sigma_{0}}{d\sqrt{\tau}dx_{F}}}
  \label{Kexp}
\end{equation}
where $d^{2}\sigma_{i}/d\sqrt{\tau}dx_{F}$ denotes the order $\alpha_{s}^{i}$
corrected cross section. Fig.~\ref{Kfactor} shows that neither $K_{1}$
nor $K_{2}$
fit the data. The second order corrected $K$-factor is closer to the data in
the small $\sqrt{\tau}$-region. It is a pity that due to the presence of the
$\Upsilon$ in the data it is difficult to compare theory with experiment in
particular in those regions of $\sqrt{\tau}$ where the $S+V$ approximation is
supposed to work.\\
Finally we also made a comparison with the data obtained by the E772
experiment for the reaction $p + \mbox{}^{2}H \rightarrow \mu^{+}\mu^{-}
+"X"$ carried out at $\sqrt{S}=38.8\,\mbox{GeV}/c$. The main goal of this
experiment was to find a charge asymmetry in the sea-quark densities of the
nucleon i.e. $\bar{u}(x) \neq \bar{d}(x)$. Here we are also interested whether
the data obtained for $m^{3}d^{2}\sigma/dmdx_{F}$ are in agreement with the
order $\alpha_{s}^{2}$ corrected DY cross section. In fig.~\ref{deut_dmin}
we have plotted the data for $m=8.15\,\mbox{GeV}/c^2$ and compared them with
the
predictions given by the Born, the order $\alpha_{s}$ corrected and the order
$\alpha_{s}^{2}$ corrected cross section. The figure shows that the order
$\alpha_{s}^{2}$ corrections are needed to bring theory into agreement with
the data. Notice that at this $m$-value one obtains
$\sqrt{\tau}=0.21$ which is
quite small for the $S+V$ approximation so that the result has to be
interpreted with care.
In the next figure (fig.~\ref{deut_asymm}) we study the effect of the higher
order QCD
corrections on the suppression of the cross section near $x_{F}=0.0$ which is
caused by the difference between the up-sea and down-sea quark densities.
Notice that the $p\,p$ reaction is symmetric whereas the $p\,n$ reaction is
asymmetric around $x_{F}=0.0$ irrespective whether there is charge asymmetry
or not. Therefore the $p\,n$ reaction leads to an $x_{F}$ asymmetry even for
isoscalar targets like $\mbox{}^{2}H$. In fig.~\ref{deut_asymm}
we have presented the
order $\alpha_{s}^{2}$ corrected cross section for three different parton
density sets for the nucleon.
They are given by MRS(S0) and MRS(D0) where the former has a
symmetric sea ($\bar{u}(x)=\bar{d}(x)$) whereas the latter contains an
asymmetric sea ($\bar{u}(x)\neq\bar{d}(x)$) parametrization. For comparison we
have also shown MRS(D-) which only differs from MRS(D0) that the gluon and
sea densities have a much steeper small $x$-behavior (lipatov-pomeron) than
the ones given by MRS(D0) and MRS(S0) (non perturbative pomeron).
Fig.~\ref{deut_asymm} reveals that there is hardly any suppression of
the cross section for $x_{F} < 0$ while going from the symmetric sea
(MRS(S0)) to the asymmetric sea (MRS(D0)) parametrization so that both parton
density sets are in agreement with the data.\\
If other parton densities are used like those discussed in \cite{r22} the
suppression for $x_{F} < 0$ can be much larger. For the MRS-set it appears
that a change in the small $x$-behavior of the parton densities leads to a
larger suppression of the cross section (compare MRS(D0) with MRS(D-)) than
the introduction of a charge asymmetry in the sea-quarks (MRS(S0) versus
MRS(D0)).\\
In addition to the calculations performed in the DIS-scheme we have also
presented in figs.~\ref{dsdm_pbW}-\ref{deut_dmin} the order $\alpha_{s}^{2}$
corrected cross section computed in the $\overline{\mbox{MS}}$-scheme.
Although the latter is an improvement with respect to the order $\alpha_{s}$
corrected result it is smaller than the cross section computed in the
DIS-scheme except when $x_{F}$ is large. This is not
surprising because figs.~\ref{ana_ms}-\ref{mcg_ms} already indicate that the
approximation underestimates the exact cross section in the case of the
$\overline{\mbox{MS}}$-scheme.\\
Summarizing the content of this work we can conclude that up to the order
$\alpha_{s}$ level the soft plus virtual gluon contribution gives a fairly
good approximation of the exact DY cross section $d^{2}\sigma/dmdx_{F}$.
Therefore we expect that this approximation will also work for the
$\alpha_{s}^{2}$ correction as long as the cross section is computed at fixed
target energies and for $\sqrt{\tau} > 0.3$. In this $\tau$-region we expect
that all other partonic subprocesses are suppressed due to the reduction in
phase space. This expectation is corroborated by a thorough analysis of the
second order contribution to $d\sigma/dm$ for which the exact coefficient
function is known. Because of the missing pieces in the order $\alpha_{s}^{2}$
contribution to the coefficient function corresponding to the cross section
$d^{2}\sigma/dmdx_{F}$ and the absence of the next-to-next-to-leading order
parton densities we have to rely on the order $\alpha_{s}^{2}$ soft plus
virtual gluon approximation to make a comparison with the data. Using this
approach we can show that a part of the discrepancy between the data and the
order $\alpha_{s}$ corrected cross section can be attributed to the higher
order
soft plus virtual gluon contributions.
\newpage
\appendix
\section{}
In this appendix we will present the order $\alpha_{s}$ contributions to the
coefficient functions corresponding to $d^{2}\sigma/dQ^{2}dx_{F}$ coming from
the partonic subprocesses in (\ref{qqvg}) and (\ref{gqvq}).
Although these processes have
been calculated in the DIS-scheme in \cite{r16,r17} (see also \cite{r26}) and
the $\overline{\mbox{MS}}$-scheme \cite{r27} we have some different
definitions for the distributions and we have a small disagreement with the
coefficient function for the $qg$ subprocess in \cite{r27}. Moreover we want
to give a clear definition for the soft plus virtual ($S+V$) gluon part of
the coefficient function corresponding to the $q\bar{q}$ subprocess.\\
We have recalculated the double differential cross section
$d^{2}\sigma/dQ^{2}dx_{F}$ for the partonic subprocesses (\ref{qqvg})
and (\ref{gqvq}). After
performing the mass factorization in the $\overline{\mbox{MS}}$-scheme
the coefficients $\Delta_{ij}^{(1)}$ (see the definition in (\ref{def_deltas}))
read as follows
\begin{eqnarray}
\lefteqn{\Delta_{q\bar{q}}^{(1)}(t_{1},t_{2},x_{1},x_{2},Q^{2},\mu^{2}) =
    C_{F}\frac{\delta(t_{1}-x_{1})\delta(t_{2}-x_{2})}{x_{1}+x_{2}}
    \left[\vphantom{\frac{A}{A}}\right.} \nonumber \\[2ex]
  && \left.\left\{ 4\ln\frac{(1-x_{1})(1-x_{2})}{x_{1}x_{2}} + 6\right\}
     \ln\frac{Q^{2}}{\mu^{2}} - 16 + 12\zeta(2)
     + 2\ln^{2}\frac{(1-x_{1})(1-x_{2})}{x_{1}x_{2}}
     \vphantom{\frac{A}{A}}\right]
  \nonumber \\[2ex]
  && + C_{F}\frac{\delta(t_{1}-x_{1})}{x_{1}+x_{2}}\left[
     \left\{\left(\frac{4}{t_{2}-x_{2}}\right)_{+}-2\frac{t_{2}+x_{2}}
     {t_{2}^{2}}\right\}\ln\frac{Q^{2}}{\mu^{2}}
     +\left(\frac{4}{t_{2}-x_{2}}\right)_{+}\ln\frac{1-x_{1}}{x_{1}}
     \right. \nonumber \\[2ex]
  && + 4\left(\frac{\ln(t_{2}/x_{2}-1)}{t_{2}-x_{2}}\right)_{+}
     + \frac{4}{t_{2}-x_{2}}\ln\frac{x_{1}+x_{2}}{x_{1}+t_{2}}
     +2\frac{t_{2}-x_{2}}{t_{2}^{2}}
  \nonumber \\[2ex]
  && \left. - 2\frac{t_{2}+x_{2}}{t_{2}^{2}}\ln\frac{(x_{1}+x_{2})(1-x_{1})
     (t_{2}-x_{2})}{x_{1}x_{2}(t_{2}+x_{1})}\right]
     + [ t_{1} \leftrightarrow t_{2}, x_{1} \leftrightarrow x_{2} ]
  \nonumber \\[2ex]
  && + C_{F}\frac{1}{x_{1}+x_{2}}\left[
     \frac{4}{(t_{1}-x_{1})_{+}(t_{2}-x_{2})_{+}}
     -2\frac{t_{2}+x_{2}}{t_{2}^{2}}\left(\frac{1}{t_{1}-x_{1}}\right)_{+}
  \right. \nonumber \\[2ex]
  && -2\frac{t_{1}+x_{1}}{t_{1}^{2}}\left(\frac{1}{t_{2}-x_{2}}\right)_{+}
     -\frac{4}{(t_{2}+x_{1})(t_{1}+x_{2})}
     +\frac{2}{t_{2}(t_{1}+x_{2})}+\frac{2}{t_{1}(t_{2}+x_{1})}
  \nonumber \\[2ex]
  && \left.
     -\frac{2x_{1}}{t_{2}^{2}(t_{1}+x_{2})}
     -\frac{2x_{2}}{t_{1}^{2}(t_{2}+x_{1})}
     +\frac{2(x_{1}+x_{2})(t_{1}^{2}+t_{2}^{2})}{t_{1}^{2}t_{2}^{2}
     (t_{1}+t_{2})}\right]
  \label{d_qqb_1}
\end{eqnarray}
where the color factor $C_{F}$ is given by $C_{F}=(N^{2}-1)/2N$ (QCD :
$N=3$). In this appendix and in the next one the distributions indicated by a
plus sign in the denominator are defined as
\begin{equation}
  \int_{x_{k}}^{1}dt_{k}\left(\frac{\ln^{i}(t_{k}/x_{k}-1)}{t_{k}-x_{k}}
    \right)_{+}f(t_{k}) = \int_{x_{k}}^{1}dt_{k}\frac{\ln^{i}
    (t_{k}/x_{k}-1)}{t_{k}-x_{k}}(f(t_{k})-f(x_{k}))
  \label{d_dist_1}
\end{equation}
\begin{eqnarray}
\lefteqn{\int_{x_{1}}^{1}dt_{1}\int_{x_{2}}^{1}dt_{2}
  \left(\frac{\ln^{i}(t_{1}/x_{1}-1)}{t_{1}-x_{1}}\right)_{+}
  \left(\frac{\ln^{j}(t_{2}/x_{2}-1)}{t_{2}-x_{2}}\right)_{+}
  f(t_{1},t_{2})} \nonumber \\[2ex]
  && =\int_{x_{1}}^{1}dt_{1}\int_{x_{2}}^{1}dt_{2}\frac{\ln^{i}
     (t_{1}/x_{1}-1)}{t_{1}-x_{1}}\frac{\ln^{j}(t_{2}/x_{2}-1)}
     {t_{2}-x_{2}}(f(t_{1},t_{2})-f(x_{1},t_{2}).
  \nonumber \\[2ex]
  && \hspace{5cm} -f(t_{1},x_{2}) +f(x_{1},x_{2}))
\end{eqnarray}
Expression (\ref{d_qqb_1}) for $\Delta_{q\bar{q}}^{(1)}$ is in agreement
with eq. (A.4) in \cite{r27}. Notice that the authors in \cite{r27} give
a different definition for the distributions.
This leads to a difference between (\ref{d_dist_1}) and eq. (A.12) in
\cite{r27} which equals
\begin{equation}
  \int_{x_{k}}^{1}dt_{k}\frac{\ln t_{k}/x_{k}}{t_{k}-x_{k}}f(x_{k})
    = f(x_{k})\left[\frac{1}{2}\ln^{2}x_{k} + \mbox{Li}_{2}(1-x_{k})\right].
  \label{dilog_diff}
\end{equation}
where the dilogarithmic function $\mbox{Li}_{2}(x)$ is defined by
\begin{equation}
  \mbox{Li}_{2}(x) = -\int_{0}^{x} \frac{dt}{t}\ln(1-t)
\end{equation}
The expression between the square brackets in (\ref{dilog_diff}),
multiplied by two, has to be added to the coefficient of the
$\delta(t_{1}-x_{1})\delta(t_{2}-x_{2})$ term in eq. (A.4) of \cite{r27}
so that one obtains the same result as we have in (\ref{d_qqb_1}) above.\\
The soft plus virtual gluon part of $\Delta_{q\bar{q}}^{(1)}$ is defined by
isolating the double singular terms in (\ref{d_qqb_1}) of the types
$\delta(t_{1}-x_{1})\delta(t_{2}-x_{2})$, $\delta(t_{1}-x_{1})\left(
\frac{1}{t_{2}-x_{2}}\right)_{+}$, $\delta(t_{2}-x_{2})\left(
\frac{1}{t_{1}-x_{1}}\right)_{+}$ and $\left(\frac{1}{t_{1}-x_{1}}\right)_{+}
\left(\frac{1}{t_{2}-x_{2}}\right)_{+}$. Hence we obtain
\begin{eqnarray}
\lefteqn{\Delta_{q\bar{q}}^{S+V,(1)}(t_{1},t_{2},x_{1},x_{2},Q^{2},\mu^{2})
  = C_{F}\frac{\delta(t_{1}-x_{1})\delta(t_{2}-x_{2})}{x_{1}+x_{2}}
  \left[\vphantom{\frac{A}{A}}\right.}
  \nonumber \\[2ex]
  && \left.\left\{4\ln\frac{(1-x_{1})(1-x_{2})}{x_{1}x_{2}} + 6\right\}
     \ln\frac{Q^{2}}{\mu^{2}} - 16 + 12\zeta(2) + 2\ln^{2}
     \frac{(1-x_{1})(1-x_{2})}{x_{1}x_{2}}\right]
  \nonumber \\[2ex]
  && +C_{F}\frac{\delta(t_{1}-x_{1})}{x_{1}+x_{2}}\left[
     \left(\frac{4}{t_{2}-x_{2}}\right)_{+}\ln\frac{Q^{2}}{\mu^{2}}
     +\left(\frac{4}{t_{2}-x_{2}}\right)_{+}\ln\frac{1-x_{1}}{x_{1}}
     \right. \nonumber \\[2ex]
  && \left.+ 4\left(\frac{\ln(t_{2}/x_{2}-1)}{t_{2}-x_{2}}\right)_{+}\right]
     +[t_{1}\leftrightarrow t_{2}, x_{1}\leftrightarrow x_{2}]
  \nonumber \\[2ex]
  && + C_{F}\frac{1}{x_{1}+x_{2}}\left[\frac{4}
     {(t_{1}-x_{1})_{+}(t_{2}-x_{2})_{+}}\right]
  \label{d_qqb_sv}
\end{eqnarray}
where we have taken the residues at $t_{k} = x_{k}$.\\
For the $gq$ subprocess we obtain the coefficient function
\begin{eqnarray}
\lefteqn{\Delta_{gq}^{(1)}(t_{1},t_{2},x_{1},x_{2},Q^{2},\mu^{2}) =
  T_{f}\frac{\delta(t_{2}-x_{2})}{x_{1}+x_{2}}\left[
  \left\{2\frac{x_{1}^{2}+(t_{1}-x_{1})^{2}}{t_{1}^{3}}
  \right\}\ln\frac{Q^{2}}{\mu^{2}}\right.}
  \nonumber \\[2ex]
  && \left.+ 2\frac{x_{1}^{2}+(t_{1}-x_{1})^{2}}{t_{1}^{3}}\ln
     \frac{(x_{1}+x_{2})(1-x_{2})(t_{1}-x_{1})}{x_{1}x_{2}(t_{1}+x_{2})}
     + \frac{4x_{1}(t_{1}-x_{1})}{t_{1}^{3}}\right]
  \nonumber \\[2ex]
  && + T_{f}\frac{1}{x_{1}+x_{2}}\left[2\frac{x_{1}^{2}+(t_{1}-x_{1})^{2}}
     {t_{1}^{3}}\left(\frac{1}{t_{2}-x_{2}}\right)_{+} - 2\frac
     {x_{2}^{2}+(t_{1}+x_{2})^{2}}{t_{1}^{3}(t_{2}+x_{1})}\right.
     +4\frac{x_{1}+x_{2}}{t_{1}^{2}t_{2}}
  \nonumber \\[2ex]
  && - 4\frac{x_{1}x_{2}(x_{1}+x_{2})}
     {t_{1}^{3}t_{2}^{2}} -4\frac{x_{1}^{2}-x_{2}^{2}}{t_{1}^{3}t_{2}}
     + 2\frac{(x_{1}+x_{2})(t_{2}-x_{2})(t_{2}+x_{1})}
     {t_{1}t_{2}^{2}(t_{1}+t_{2})^{2}}
  \nonumber \\[2ex]
  && \left.+ 4\frac{x_{1}x_{2}(x_{1}+x_{2})}
     {t_{1}^{2}t_{2}^{2}(t_{1}+t_{2})}\right]
  \label{d_gq_1}
\end{eqnarray}
and
\begin{equation}
  \Delta_{qg}^{(1)}(t_{1},t_{2},x_{1},x_{2},Q^{2},\mu^{2}) =
    \Delta_{gq}^{(1)}(t_{2},t_{1},x_{2},x_{1},Q^{2},\mu^{2})
  \label{d_qg_1}
\end{equation}
where $T_{f} = 1/2$.\\
There is a discrepancy between our answer in (\ref{d_gq_1}) and the one
given in eq. (A.8) of \cite{r27}. The difference between their result and
ours equals $2\frac{x_{1}^{2}+(t_{1}-x_{1})^{2}}{t_{1}^{3}}$. This
discrepancy can be attributed to the procedure that in $n$-dimensional
regularization before mass factorization the cross section with one gluon in
the initial state has to be divided by $n-2$ in order to average over the
initial gluon polarizations. Only in this case one can combine the coefficient
functions with the parton densities of which the scale evolution is
determined by the two-loop anomalous dimensions (or Altarelli-Parisi
splitting functions) calculated in the literature (see e.g. \cite{r31}). The
expression in eq. (A.8) of \cite{r27} can be only obtained if the
polarization average factor is a $1/2$ instead of $1/(n-2)$.
In the latter case one has
to modify the two-loop anomalous dimensions via a finite renormalization.
However the MRS parton densities in \cite{r28} were constructed using the
anomalous dimensions in \cite{r31} so that one has to divide the parton cross
section by $n-2$ and not by $2$. The choice of the polarization average
factor shows
up again when we want to present the coefficient functions in the DIS-scheme.
The results in the DIS-scheme are obtained by performing a finite mass
factorization. The coefficient functions in the two schemes are related by
\begin{eqnarray}
\lefteqn{\left.\Delta_{ij}(t_{1},t_{2},x_{1},x_{2},Q^{2},\mu^{2})
  \right|_{\mbox{DIS}} = \sum_{k,l}\int_{x_{1}}^{1}du_{1}
  \int_{x_{2}}^{1}du_{2}\Gamma_{ki}\left(\frac{u_{1}}{t_{1}}
  \right)\Gamma_{lj}\left(\frac{u_{2}}{t_{2}}\right)}
  \nonumber \\[2ex]
  && \hspace{3cm}\left.
     \Delta_{kl}(u_{1},u_{2},x_{1},x_{2},Q^{2},\mu^{2})\right|
     _{\overline{\mbox{MS}}}.
\end{eqnarray}
Up to order $\alpha_{s}$, $\Gamma_{qq}(x)$ and $\Gamma_{qg}(x)$ are given by
\begin{eqnarray}
  \Gamma_{qq}(x) &=& \delta(1-x) + \frac{\alpha_{s}}{4\pi}C_{F}\left[
    4\left(\frac{\ln(1-x)}{1-x}\right)_{+} - 2 (1+x)\ln(1-x)+6+4x\right.
  \nonumber \\[2ex]
  && \left.-2\frac{1+x^{2}}{1-x}\ln x - \left(\frac{3}{1-x}\right)_{+}
     + \delta(1-x)(-9-4\zeta(2))\right]
  \label{gamma_qq}
\end{eqnarray}
\begin{equation}
  \Gamma_{qg}(x) = \frac{\alpha_{s}}{4\pi}T_{f}\left[
    2\left\{x^{2}+(1-x)^2\right\}\ln\frac{1-x}{x}
    + 16x(1-x) - 2 \right].
  \label{gamma_qg}
\end{equation}
Expressions (\ref{gamma_qq}) and (\ref{gamma_qg}) are in agreement with
$C_{F,2}^{(1)}$ and
$C_{G,2}^{(1)}$ in appendix I of \cite{r32}. Notice that the authors in
\cite{r27} used a $\Gamma_{qg}(x)$ where $16x(1-x)-2$ is replaced by
$12x(1-x)$ which is obtained when the gluon polarization average factor
is taken to be $1/2$ instead of $1/(n-2)$. See the discussion above.\\
The coefficient functions in the DIS-scheme read
\begin{eqnarray}
\lefteqn{\Delta_{q\bar{q}}^{(1)}(t_{1},t_{2},x_{1},x_{2},Q^{2},\mu^{2}) =
  C_{F}\frac{\delta(t_{1}-x_{1})\delta(t_{2}-x_{2})}{x_{1}+x_{2}}\left[
  4\ln\frac{1-x_{1}}{x_{1}}\ln\frac{1-x_{2}}{x_{2}}
  \vphantom{\frac{A}{A}}\right.}
  \nonumber \\[2ex]
  && \left.+\left\{4\ln\frac{(1-x_{1})(1-x_{2})}{x_{1}x_{2}} + 6\right\}
     \ln\frac{Q^{2}}{\mu^{2}} + 2 + 20\zeta(2)
     + 3\ln\frac{(1-x_{1})(1-x_{2})}{x_{1}x_{2}}\right]
  \nonumber \\[2ex]
  && +C_{F}\frac{\delta(t_{1}-x_{1})}{x_{1}+x_{2}}\left[\left\{
     \left(\frac{4}{t_{2}-x_{2}}\right)_{+} - 2\frac{t_{2}+x_{2}}{t_{2}^{2}}
     \right\}\ln\frac{Q^{2}}{\mu^{2}}\right.
  \nonumber \\[2ex]
  && + \left(4\ln\frac{1-x_{1}}{x_{1}}
     + 3\right)\left(\frac{1}{t_{2}-x_{2}}\right)_{+} +
     \frac{4}{t_{2}-x_{2}}\ln\frac{x_{1}+x_{2}}{x_{1}+t_{2}}
  \nonumber \\[2ex]
  && \left. - 2\frac
     {t_{2}+x_{2}}{t_{2}^{2}}\ln\frac{(x_{1}+x_{2})(1-x_{1})}
     {x_{1}(t_{2}+x_{1})} - \frac{4}{t_{2}} - 6\frac{x_{2}}{t_{2}^{2}}
     \right] + [t_{1}\leftrightarrow t_{2}, x_{1}\leftrightarrow x_{2}]
  \nonumber \\[2ex]
  && + C_{F}\frac{1}{x_{1}+x_{2}}\left[
     \frac{4}{(t_{1}-x_{1})_{+}(t_{2}-x_{2})_{+}} -2\frac{t_{2}+x_{2}}
     {t_{2}^{2}}\left(\frac{1}{t_{1}-x_{1}}\right)_{+}\right.
  \nonumber \\[2ex]
  && -2\frac{t_{1}+x_{1}}{t_{2}^{2}}\left(\frac{1}{t_{2}-x_{2}}\right)_{+}
     -\frac{4}{(t_{2}+x_{1})(t_{1}+x_{2})} + \frac{2}{t_{2}(t_{1}+x_{2})}
     +\frac{2}{t_{1}(t_{2}+x_{1})}
  \nonumber \\[2ex]
  && \left.-2\frac{x_{1}}{t_{2}^{2}(t_{1}+x_{2})} -2\frac{x_{2}}
     {t_{1}^{2}(t_{2}+x_{1})} + 2\frac{(x_{1}+x_{2})(t_{1}^{2}+t_{2}^{2})}
     {t_{1}^{2}t_{2}^{2}(t_{1}+t_{2})}\right].
\end{eqnarray}
The soft plus virtual gluon part is obtained in the same way as discussed in
the case of the $\overline{\mbox{MS}}$-scheme
\begin{eqnarray}
\lefteqn{\Delta_{q\bar{q}}^{S+V,(1)}(t_{1},t_{2},x_{1},x_{2},Q^{2},\mu^{2})
  = C_{F}\frac{\delta(t_{1}-x_{1})\delta(t_{2}-x_{2})}{x_{1}+x_{2}}
  \left[4\ln\frac{1-x_{1}}{x_{1}}\ln\frac{1-x_{2}}{x_{2}}
  \vphantom{\frac{A}{A}}\right.}
  \nonumber \\[2ex]
  && \left.\left\{4\ln\frac{(1-x_{1})(1-x_{2})}{x_{1}x_{2}} + 6\right\}
     \ln\frac{Q^{2}}{\mu^{2}} + 2 + 20\zeta(2)
     + 3\ln\frac{(1-x_{1})(1-x_{2})}{x_{1}x_{2}}\right]
  \nonumber \\[2ex]
  && + C_{F}\frac{\delta(t_{1}-x_{1})}{x_{1}+x_{2}}\left[\left\{
     \left(\frac{4}{t_{2}-x_{2}}\right)_{+}\right\}\ln\frac{Q^{2}}{\mu^{2}}
     \right.
  \nonumber \\[2ex]
  && \left.+\left(4\ln\frac{1-x_{1}}{x_{1}} + 3\right)\left(
     \frac{1}{t_{2}-x_{2}}\right)_{+}\right]
     + [ t_{1}\leftrightarrow t_{2}, x_{1}\leftrightarrow x_{2}]
  \nonumber \\[2ex]
  && + C_{F}\frac{1}{x_{1}+x_{2}}\left[\frac{4}{(t_{1}-x_{1})_{+}
     (t_{2}-x_{2})_{+}}\right].
  \label{d_qqb_sv_dis}
\end{eqnarray}
The coefficient function for the subprocess with the gluon in the initial
state becomes
\begin{eqnarray}
\lefteqn{\Delta_{gq}^{(1)}(t_{1},t_{2},x_{1},x_{2},Q^{2},\mu^{2}) =
  T_{f}\frac{\delta(t_{2}-x_{2})}{x_{1}+x_{2}}\left[
  \left\{2\frac{x_{1}^{2}+(t_{1}-x_{1})^{2}}{t_{1}^{3}}\right\}
  \ln\frac{Q^{2}}{\mu^{2}}\right.}
  \nonumber \\[2ex]
  && \left.+ 2\frac{x_{1}^{2}+(t_{1}-x_{1})^{2}}{t_{1}^{3}}\ln\frac
     {(x_{1}+x_{2})(1-x_{2})}{x_{2}(t_{1}+x_{2})} + \frac{2}{t_{1}}
     - 12\frac{x_{1}(t_{1}-x_{1})}{t_{1}^{3}}\right]
  \nonumber \\[2ex]
  && + T_{f}\frac{1}{x_{1}+x_{2}}\left[ 2\frac{x_{1}^{2}+(t_{1}-x_{1})^{2}}
     {t_{1}^{3}}\left(\frac{1}{t_{2}-x_{2}}\right)_{+} -
     2\frac{x_{2}^{2}+(t_{1}+x_{2})^{2}}{t_{1}^{3}(t_{2}+x_{1})}
     + 4\frac{x_{1}+x_{2}}{t_{1}^{2}t_{2}}\right.
  \nonumber \\[2ex]
  && -4\frac{x_{1}x_{2}(x_{1}+x_{2})}{t_{1}^{3}t_{2}^{2}}
     -4\frac{x_{1}^{2}-x_{2}^{2}}{t_{1}^{3}t_{2}} + 2\frac
     {(x_{1}+x_{2})(t_{2}-x_{2})(t_{2}+x_{1})}{t_{1}t_{2}^{2}
     (t_{1}+t_{2})^{2}}
  \nonumber \\[2ex]
  && \left.+ 4\frac{x_{1}x_{2}(x_{1}+x_{2})}
     {t_{1}^{2}t_{2}^{2}(t_{1}+t_{2})}\right]
\end{eqnarray}
where $\Delta_{qg}^{(1)}$ is related to $\Delta_{gq}^{(1)}$ via relation
(\ref{d_qg_1}).\\
We have explicitly checked that if the above coefficient functions are
inserted in (\ref{diff_cross}) and the integrals over $x_{F}$ are performed
according to (\ref{dsdq}) one gets the same answer as given by $d\sigma/dm$
(\ref{dsdq_2}) with the coefficient functions obtained from
\cite{r23,r24}. The coefficient functions for $d^{2}\sigma/dQ^{2}dy$
have not been explicitly listed here but are present in our computer program
DIFDY. In the case of $d^{2}\sigma/dQ^{2}dy$ we
agree with the results for the $\overline{\mbox{MS}}$-scheme published in
\cite{r27} except for $1/2t_{1}$ in eq. (A.20) which has to be
replaced by $\frac{x_{1}(t_{1}-x_{1})}{t_{1}^{3}}$. This difference follows
again from taking the average over the initial gluon polarizations as discussed
for $\Delta_{gq}^{(1)}$ above. Our results for the DIS-scheme agree with those
presented in the appendix of \cite{r26}. Notice that the soft plus virtual
gluon part of $\Delta_{q\bar{q}}^{(1)}$ for $d^{2}\sigma/dQ^{2}dy$ can be
obtained from (\ref{d_qqb_sv}) and (\ref{d_qqb_sv_dis}) by multiplication
with $x_{1}+x_{2}$.
\newpage
\section{}
The order $\alpha_{s}^{2}$ contribution to the coefficient function in the
$S+V$ approximation has been calculated in \cite{r25}. Including the mass
factorization parts represented by $\ln Q^{2}/\mu^{2}$ and rewriting
the coefficient
function in a more amenable form as presented for the first order correction
in appendix A it reads in the $\overline{\mbox{MS}}$-scheme as follows
\begin{eqnarray}
\lefteqn{\Delta_{q\bar{q}}^{S+V,(2)}(t_{1},t_{2},x_{1},x_{2},Q^{2},\mu^{2})
  = \frac{\delta(t_{1}-x_{1})\delta(t_{2}-x_{2})}{x_{1}+x_{2}}\left[
  \vphantom{\frac{A}{A}}\right.}
  \nonumber \\[2ex]
  && C_{F}^{2}\left\{\left[\vphantom{\frac{A}{A}}
     18 - 8\zeta(2) + 8\left( P^{2}(x_{1})
     + P^{2}(x_{2})\right) + 24 \left(P(x_{1}) + P(x_{2})\right)
     \right.\right.
  \nonumber \\[2ex]
  && \left.+ 16 P(x_{1}) P(x_{2})\vphantom{\frac{A}{A}}\right] L_{\mu}^{2}
     +\left[-93+60\zeta(2)+80\zeta(3)+8\left(P^{3}(x_{1})+
     P^{3}(x_{2})\right)\right.
  \nonumber \\[2ex]
  && + 24\left(P^{2}(x_{1})P(x_{2}) + P(x_{1})P^{2}(x_{2})\right)
     + 24 P(x_{1})P(x_{2})
  \nonumber \\[2ex]
  && \left.+ 12\left(P^{2}(x_{1})+P^{2}(x_{2})\right) + (-64+16\zeta(2))
     (P(x_{1}) + P(x_{2}))\right] L_{\mu}
  \nonumber \\[2ex]
  && + \frac{511}{4} - 128\zeta(2)- 60\zeta(3)+\frac{304}{9}
     \zeta^{2}(2) + 2\left(P^{4}(x_{1})+P^{4}(x_{2})\right)
  \nonumber \\[2ex]
  && + 8\left(P^{3}(x_{1})P(x_{2}) + P(x_{1})P^{3}(x_{2})\right)
     + 12P^{2}(x_{1})P^{2}(x_{2})
  \nonumber \\[2ex]
  && + (-32 + 8\zeta(2))\left(P^{2}(x_{1})+P^{2}(x_{2})\right)
     + (-64+16\zeta(2))P(x_{1})P(x_{2})
  \nonumber \\[2ex]
  && \left.+ 32\zeta(3)(P(x_{1})+P(x_{2}))\vphantom{\frac{A}{A}}\right\}
  \nonumber \\[2ex]
  && + C_{A}C_{F}\left\{\left[ -11 -\frac{22}{3}(P(x_{1})+P(x_{2}))
     \right] L_{\mu}^{2} + \left[\frac{193}{3} - 22\zeta(2) - 24\zeta(3)
     \right.\right.
  \nonumber \\[2ex]
  && -\frac{22}{3}\left(P^{2}(x_{1})+P^{2}(x_{2})\right)
     +\left(\frac{268}{9}-8\zeta(2)\right)(P(x_{1})+P(x_{2}))
  \nonumber \\[2ex]
  && \left.\vphantom{\frac{A}{A}}
     -\frac{44}{3}P(x_{1})P(x_{2})\right] L_{\mu}
     -\frac{1535}{12} + \frac{860}{9}\zeta(2) + \frac{172}{3}\zeta(3)
     -\frac{52}{5}\zeta^{2}(2)
  \nonumber \\[2ex]
  && - \frac{22}{9}\left( P^{3}(x_{1}) + P^{3}(x_{2})\right)
     -\frac{22}{3}\left(P^{2}(x_{1})P(x_{2}) + P(x_{1})P^{2}(x_{2})\right)
  \nonumber \\[2ex]
  && + \left(\frac{134}{9}-4\zeta(2)\right)\left(P^{2}(x_{1})
     + P^{2}(x_{2})\right)
     + \left(\frac{268}{9} - 8\zeta(2)\right)P(x_{1})P(x_{2})
  \nonumber \\[2ex]
  && \left.+ \left(-\frac{808}{27}+\frac{44}{3}\zeta(2)+28\zeta(3)\right)
     (P(x_{1})+P(x_{2}))\right\}
  \nonumber \\[2ex]
  && + n_{f}C_{F}\left\{\left[ 2 +\frac{4}{3}(P(x_{1})+P(x_{2}))
     \right] L_{\mu}^{2} + \left[-\frac{34}{3}+4\zeta(2)+\frac{4}{3}
     \left(P^{2}(x_{1})+P^{2}(x_{2})\right)\right.\right.\hspace*{-11pt}
  \nonumber \\[2ex]
  && \left.+\frac{8}{3}P(x_{1})P(x_{2}) - \frac{40}{9}(P(x_{1})+P(x_{2}))
     \right] L_{\mu} + \frac{127}{6} - \frac{152}{9}\zeta(2)
     +\frac{8}{3}\zeta(3)
  \nonumber \\[2ex]
  && +\frac{4}{9}\left(P^{3}(x_{1})+P^{3}(x_{2})\right)+
     \frac{4}{3}\left(P^{2}(x_{1})P(x_{2})+P(x_{1})P^{2}(x_{2})\right)
     -\frac{40}{9}P(x_{1})P(x_{2})
  \nonumber \\[2ex]
  && \left.\left.-\frac{20}{9}\left(P^{2}(x_{1})+P^{2}(x_{2})\right)
     +\left(\frac{112}{27}-\frac{8}{3}\zeta(2)\right)
     (P(x_{1})+P(x_{2})) \right\}\right]
  \nonumber \\[2ex]
  && + \frac{\delta(t_{1}-x_{1})}{x_{1}+x_{2}}\left[
     C_{F}^{2}\left\{\left[ \vphantom{\frac{A}{A}}
     16 (D_{1}(t_{2}) + P(x_{1})D_{0}(t_{2}))
     +24D_{0}(t_{2})\right] L_{\mu}^{2} +\left[\vphantom{\frac{A}{A}}
     24D_{2}(t_{2})\right.\right.\right.
  \nonumber \\[2ex]
  && +48P(x_{1})D_{1}(t_{2})+24P^{2}(x_{1})D_{0}(t_{2})+24(D_{1}(t_{2})
     +P(x_{1})D_{0}(t_{2}))
  \nonumber \\[2ex]
  && \left.\vphantom{\frac{a}{a}}
     +(-64+16\zeta(2))D_{0}(t_{2})\right] L_{\mu} +8D_{3}(t_{2})
     +24P(x_{1})D_{2}(t_{2})+24P^{2}(x_{1})D_{1}(t_{2})
  \nonumber \\[2ex]
  && \left.\vphantom{\frac{a}{a}}
     +8P^{3}(x_{1})D_{0}(t_{2}) + (-64+16\zeta(2))(D_{1}(t_{2})
     +P(x_{1})D_{0}(t_{2}))+32\zeta(3)D_{0}(t_{2})\right\}\hspace*{-6pt}
  \nonumber \\[2ex]
  && + C_{A}C_{F}\left\{\left[-\frac{22}{3}D_{0}(t_{2})\right]L_{\mu}^{2}
     +\left[-\frac{44}{3}D_{1}(t_{2})-\frac{44}{3}P(x_{1})D_{0}(t_{2})
     \right.\right.
  \nonumber \\[2ex]
  && \left.+\left(\frac{268}{9}-8\zeta(2)\right)D_{0}(t_{2})\right] L_{\mu}
     -\frac{22}{3}D_{2}(t_{2}) -\frac{44}{3}P(x_{1})D_{1}(t_{2})
  \nonumber \\[2ex]
  && -\frac{22}{3}P^{2}(x_{1})D_{0}(t_{2})
     +\left(\frac{268}{9}-8\zeta(2)\right)(D_{1}(t_{2}) +
     P(x_{1})D_{0}(t_{2}))
  \nonumber \\[2ex]
  && \left.+\left(-\frac{808}{27}+\frac{44}{3}\zeta(2)+28\zeta(3)\right)
     D_{0}(t_{2})\right\}
  \nonumber \\[2ex]
  && +n_{f}C_{F} \left\{\left[\frac{4}{3}D_{0}(t_{2})\right]L_{\mu}^{2}
     +\left[\frac{8}{3}(D_{1}(t_{2})+P(x_{1})D_{0}(t_{2}))
     -\frac{40}{9}D_{0}(t_{2})\right] L_{\mu}\right.
  \nonumber \\[2ex]
  && +\frac{4}{3}D_{2}(t_{2})+\frac{8}{3}P(x_{1})D_{1}(t_{2})
     +\frac{4}{3}P^{2}(x_{1})D_{0}(t_{2})-\frac{40}{9}(D_{1}(t_{2})
     +P(x_{1})D_{0}(t_{2}))
  \nonumber \\[2ex]
  && \left.\left.+\left(\frac{112}{27}-\frac{8}{3}\zeta(2)\right)
     D_{0}(t_{2}) \right\}\right]
     +[t_{1}\leftrightarrow t_{2}, x_{1}\leftrightarrow x_{2}]
  \nonumber \\[2ex]
  && +\frac{1}{x_{1}+x_{2}}\left[C_{F}^{2}\left\{\left[
     \vphantom{\frac{A}{A}}16D_{0}(t_{1})D_{0}(t_{2})\right] L_{\mu}^{2}
     +\left[\vphantom{\frac{A}{A}}48(D_{1}(t_{1})D_{0}(t_{2})
     +D_{0}(t_{1})D_{1}(t_{2}))\right.\right.\right.
  \nonumber \\[2ex]
  && \left.\vphantom{\frac{A}{A}}
     +24D_{0}(t_{1})D_{0}(t_{2})\right] L_{\mu}
     +48D_{1}(t_{1})D_{1}(t_{2})
  \nonumber \\[2ex]
  && \left.\vphantom{\frac{A}{A}}
     +24(D_{2}(t_{1})D_{0}(t_{2})+D_{0}(t_{1})D_{2}(t_{2}))
     +(-64+16\zeta(2))D_{0}(t_{1})D_{0}(t_{2})\right\}
  \nonumber \\[2ex]
  && + C_{A}C_{F}\left\{\left[-\frac{44}{3}D_{0}(t_{1})D_{0}(t_{2})\right]
     L_{\mu} -\frac{44}{3}(D_{1}(t_{1})D_{0}(t_{2})+D_{0}(t_{1})
     D_{1}(t_{2}))\right.
  \nonumber \\[2ex]
  && \left.\vphantom{\frac{A}{A}}
     +\left(\frac{268}{9}-8\zeta(2)\right)D_{0}(t_{1})D_{0}(t_{2})\right\}
  \nonumber \\[2ex]
  && +n_{f}C_{F}\left\{\left[\frac{8}{3}D_{0}(t_{1})D_{0}(t_{2})\right]
     L_{\mu} + \frac{8}{3}(D_{1}(t_{1})D_{0}(t_{2})+D_{0}(t_{1})
     D_{1}(t_{2}))\right.
  \nonumber \\[2ex]
  && \left.\left.\vphantom{\frac{A}{A}}
     -\frac{40}{9}D_{0}(t_{1})D_{0}(t_{2})\right\}\right]
  \label{d_qqb_sv_2}
\end{eqnarray}
Here the color factors are given by $C_{A}=N$, $C_{F}=(N^{2}-1)/2N$ (QCD :
$N=3$) and $n_{f}$ denotes the number of light flavors. In the above
expression we have introduced the following shorthand notations
\begin{equation}
  P^{i}(x_{k}) = \ln^{i}\left(\frac{1-x_{k}}{x_{k}}\right)
\end{equation}
\begin{equation}
  D_{i}(t_{k}) = \left(\frac{\ln^{i}\left(\frac{t_{k}}{x_{k}}-1\right)}
    {t_{k}-x_{k}}\right)_{+}
\end{equation}
\begin{equation}
  L_{\mu}^{i} = \ln^{i}\frac{Q^{2}}{\mu^{2}}
\end{equation}
In the DIS-scheme the above coefficient function becomes
\begin{eqnarray}
\lefteqn{\Delta_{q\bar{q}}^{S+V,(2)}(t_{1},t_{2},x_{1},x_{2},Q^{2},\mu^{2})
  = \frac{\delta(t_{1}-x_{1})\delta(t_{2}-x_{2})}{x_{1}+x_{2}}\left[
  \vphantom{\frac{A}{A}}\right.}
  \nonumber \\[2ex]
  && C_{F}^{2}\left\{\left[\vphantom{\frac{A}{A}}
     18 - 8\zeta(2) + 8\left( P^{2}(x_{1})
     + P^{2}(x_{2})\right) + 24 \left(P(x_{1}) + P(x_{2})\right)
     \right.\right.
  \nonumber \\[2ex]
  && \left.\vphantom{\frac{A}{A}}
     + 16 P(x_{1}) P(x_{2})\right] L_{\mu}^{2}
     +\left[\vphantom{\frac{A}{A}} 15+84\zeta(2)+48\zeta(3)
     + 12\left(P^{2}(x_{1})+P^{2}(x_{2})\right)\right.
  \nonumber \\[2ex]
  && + 16\left(P^{2}(x_{1})P(x_{2}) + P(x_{1})P^{2}(x_{2})\right)
     + 48 P(x_{1})P(x_{2})
  \nonumber \\[2ex]
  && \left.\vphantom{\frac{A}{A}}
     + ( 26+64\zeta(2))(P(x_{1}) + P(x_{2}))\right] L_{\mu}
     + 14\zeta(2)+ 72\zeta(3)+\frac{964}{5}
     \zeta^{2}(2)
  \nonumber \\[2ex]
  && +12\left(P^{2}(x_{1})P(x_{2})+P(x_{1})P^{2}(x_{2})\right)
     + 8P^{2}(x_{1})P^{2}(x_{2})
  \nonumber \\[2ex]
  && + ( 17+80\zeta(2))P(x_{1})P(x_{2})
     +\left(\frac{9}{2} - 8\zeta(2)\right)
     \left(P^{2}(x_{1})+P^{2}(x_{2})\right)
  \nonumber \\[2ex]
  && \left.+ \left(\frac{15}{2}+36\zeta(2)+24\zeta(3)\right)
     (P(x_{1})+P(x_{2}))\vphantom{\frac{A}{A}}\right\}
  \nonumber \\[2ex]
  && + C_{A}C_{F}\left\{\left[ -11 -\frac{22}{3}(P(x_{1})+P(x_{2}))
     \right] L_{\mu}^{2} + \left[\frac{193}{3} - 22\zeta(2) - 24\zeta(3)
     \right.\right.
  \nonumber \\[2ex]
  && -\frac{22}{3}\left(P^{2}(x_{1})+P^{2}(x_{2})\right)
     + \left(\frac{268}{9}-8\zeta(2)\right)(P(x_{1})+P(x_{2}))
  \nonumber \\[2ex]
  && \left.\vphantom{\frac{A}{A}}-\frac{44}{3}P(x_{1})P(x_{2})
     \right] L_{\mu}
     +\frac{215}{9} + \frac{2366}{9}\zeta(2) - 36\zeta(3)
     -\frac{194}{5}\zeta^{2}(2)
  \nonumber \\[2ex]
  && -\frac{22}{3}\left(P^{2}(x_{1})P(x_{2}) + P(x_{1})P^{2}(x_{2})\right)
     - \frac{11}{2}\left(P^{2}(x_{1}) + P^{2}(x_{2})\right)
  \nonumber \\[2ex]
  && \left.+ \left(\frac{57}{2}-12\zeta(3)\right)(P(x_{1})+P(x_{2}))
     + \left(\frac{268}{9} - 8\zeta(2)\right)P(x_{1})P(x_{2})
     \right\}
  \nonumber \\[2ex]
  && + n_{f}C_{F}\left\{\left[ 2 +\frac{4}{3}(P(x_{1})+P(x_{2}))
     \right] L_{\mu}^{2} + \left[-\frac{34}{3}+4\zeta(2)\right.\right.
  \nonumber \\[2ex]
  && \left.+\frac{4}{3}\left(P^{2}(x_{1})+P^{2}(x_{2})\right)
     +\frac{8}{3}P(x_{1})P(x_{2}) - \frac{40}{9}(P(x_{1})+P(x_{2}))
     \right] L_{\mu} - \frac{38}{9}
  \nonumber \\[2ex]
  && - \frac{380}{9}\zeta(2)
     -\frac{40}{9}P(x_{1})P(x_{2})
     +\frac{4}{3}\left(P^{2}(x_{1})P(x_{2})+P(x_{1})P^{2}(x_{2})\right)
  \nonumber \\[2ex]
  && \left.\left.\vphantom{\frac{A}{A}}
     +\left(P^{2}(x_{1})+P^{2}(x_{2})\right)
     -5(P(x_{1})+P(x_{2})) \right\}\right]
  \nonumber \\[2ex]
  && + \frac{\delta(t_{1}-x_{1})}{x_{1}+x_{2}}\left[
     C_{F}^{2}\left\{\left[ \vphantom{\frac{a}{a}}
     16 (D_{1}(t_{2}) + P(x_{1})D_{0}(t_{2}))
     +24D_{0}(t_{2})\right] L_{\mu}^{2}
     \right.\right.
  \nonumber \\[2ex]
  && +\left[\vphantom{\frac{a}{a}}
     32P(x_{1})D_{1}(t_{2})+16P^{2}(x_{1})D_{0}(t_{2})+24D_{1}(t_{2})
     +48P(x_{1})D_{0}(t_{2})\right.
  \nonumber \\[2ex]
  && \left.\vphantom{\frac{a}{a}}
     +( 26+64\zeta(2))D_{0}(t_{2})\right] L_{\mu}
     +16P^{2}(x_{1})D_{1}(t_{2})
     +(9-16\zeta(2))D_{1}(t_{2})
  \nonumber \\[2ex]
  && +24P(x_{1})D_{1}(t_{2})+12P^{2}(x_{1})D_{0}(t_{2})
     +(17+80\zeta(2))P(x_{1})D_{0}(t_{2}))
  \nonumber \\[2ex]
  && \left.\vphantom{\frac{a}{a}}
     +\left(\frac{15}{2}+36\zeta(2)+24\zeta(3)\right)D_{0}(t_{2})\right\}
  \nonumber \\[2ex]
  && + C_{A}C_{F}\left\{\left[-\frac{22}{3}D_{0}(t_{2})\right]L_{\mu}^{2}
     +\left[-\frac{44}{3}D_{1}(t_{2})-\frac{44}{3}P(x_{1})D_{0}(t_{2})
     \right.\right.
  \nonumber \\[2ex]
  && \left.+\left(\frac{268}{9}-8\zeta(2)\right)D_{0}(t_{2})\right] L_{\mu}
     -\frac{44}{3}P(x_{1})D_{1}(t_{2})
     -\frac{22}{3}P^{2}(x_{1})D_{0}(t_{2})
  \nonumber \\[2ex]
  && \left.-11D_{1}(t_{2})
     +\left(\frac{268}{9}-8\zeta(2)\right)P(x_{1})D_{0}(t_{2})
     +\left(\frac{57}{2}-12\zeta(3)\right)D_{0}(t_{2})\right\}
  \nonumber \\[2ex]
  && +n_{f}C_{F} \left\{\left[\frac{4}{3}D_{0}(t_{2})\right]L_{\mu}^{2}
     +\left[\frac{8}{3}(D_{1}(t_{2})+P(x_{1})D_{0}(t_{2}))
     -\frac{40}{9}D_{0}(t_{2})\right] L_{\mu}\right.
  \nonumber \\[2ex]
  && +\frac{8}{3}P(x_{1})D_{1}(t_{2})
     +\frac{4}{3}P^{2}(x_{1})D_{0}(t_{2})+2D_{1}(t_{2})
     -\frac{40}{9}P(x_{1})D_{0}(t_{2})
  \nonumber \\[2ex]
  && \left.\left.\vphantom{\frac{A}{A}}
     -5D_{0}(t_{2}) \right\}\right]
     + [t_{1} \leftrightarrow t_{2}, x_{1} \leftrightarrow x_{2}]
  \nonumber \\[2ex]
  && +\frac{1}{x_{1}+x_{2}}\left[C_{F}^{2}\left\{\left[
     \vphantom{\frac{A}{A}}16D_{0}(t_{1})D_{0}(t_{2})\right] L_{\mu}^{2}
     +\left[\vphantom{\frac{A}{A}}32(D_{1}(t_{1})D_{0}(t_{2})
     +D_{0}(t_{1})D_{1}(t_{2}))\right.\right.\right.
  \nonumber \\[2ex]
  && \left.\vphantom{\frac{A}{A}}
     +48D_{0}(t_{1})D_{0}(t_{2})\right] L_{\mu}
     +32D_{1}(t_{1})D_{1}(t_{2})
  \nonumber \\[2ex]
  && \left.\vphantom{\frac{A}{A}}
     +24(D_{1}(t_{1})D_{0}(t_{2})+D_{0}(t_{1})D_{1}(t_{2}))
     +(17+80\zeta(2))D_{0}(t_{1})D_{0}(t_{2})\right\}
  \nonumber \\[2ex]
  && + C_{A}C_{F}\left\{\left[-\frac{44}{3}D_{0}(t_{1})D_{0}(t_{2})\right]
     L_{\mu} -\frac{44}{3}(D_{1}(t_{1})D_{0}(t_{2})+D_{0}(t_{1})
     D_{1}(t_{2}))\right.
  \nonumber \\[2ex]
  && \left.\vphantom{\frac{A}{A}}
     +\left(\frac{268}{9}-8\zeta(2)\right)D_{0}(t_{1})D_{0}(t_{2})\right\}
  \nonumber \\[2ex]
  && +n_{f}C_{F}\left\{\left[\frac{8}{3}D_{0}(t_{1})D_{0}(t_{2})\right]
     L_{\mu} + \frac{8}{3}(D_{1}(t_{1})D_{0}(t_{2})+D_{0}(t_{1})
     D_{1}(t_{2}))\right.
  \nonumber \\[2ex]
  && \left.\left.\vphantom{\frac{A}{A}}
     -\frac{40}{9}D_{0}(t_{1})D_{0}(t_{2})\right\}\right].
  \label{d_qqb_sv_2_dis}
\end{eqnarray}
If one chooses the renormalization scale $\mu_{R}$ unequal to the mass
factorization scale $\mu$ one has to add the following term to the expressions
in (\ref{d_qqb_sv_2}) and (\ref{d_qqb_sv_2_dis})
\begin{equation}
  \beta_{0}\,\frac{\alpha_{s}(\mu_{R}^{2})}{4\pi}\,
    \ln\frac{\mu_{R}^{2}}{\mu^{2}}\,
    \Delta_{q\bar{q}}^{S+V,(1)}(t_{1},t_{2},x_{1},x_{2},Q^{2},\mu^{2})
  \label{r_neq_mu}
\end{equation}
where $\beta_{0}$ is the lowest order coefficient in the $\beta$-function
given by
\begin{equation}
   \beta_{0}\ = \frac{11}{3} C_{A} - \frac{2}{3} n_{f}
\end{equation}
and $\Delta_{qq}^{S+V,(1)}$ can be found in (\ref{d_qqb_1}) for the
$\overline{\mbox{MS}}$-scheme and in (\ref{d_qqb_sv_dis}) for the DIS-scheme.\\
The coefficient functions for the cross section $d^{2}\sigma/dQ^{2}dy$ can be
very easily derived from the above expression by multiplying the coefficient
functions in (\ref{d_qqb_sv_2}),(\ref{d_qqb_sv_2_dis}) and
(\ref{r_neq_mu}) by the factor $x_{1}+x_{2}$.

\begin{figure}
  \epsfysize=450pt
  \epsffile{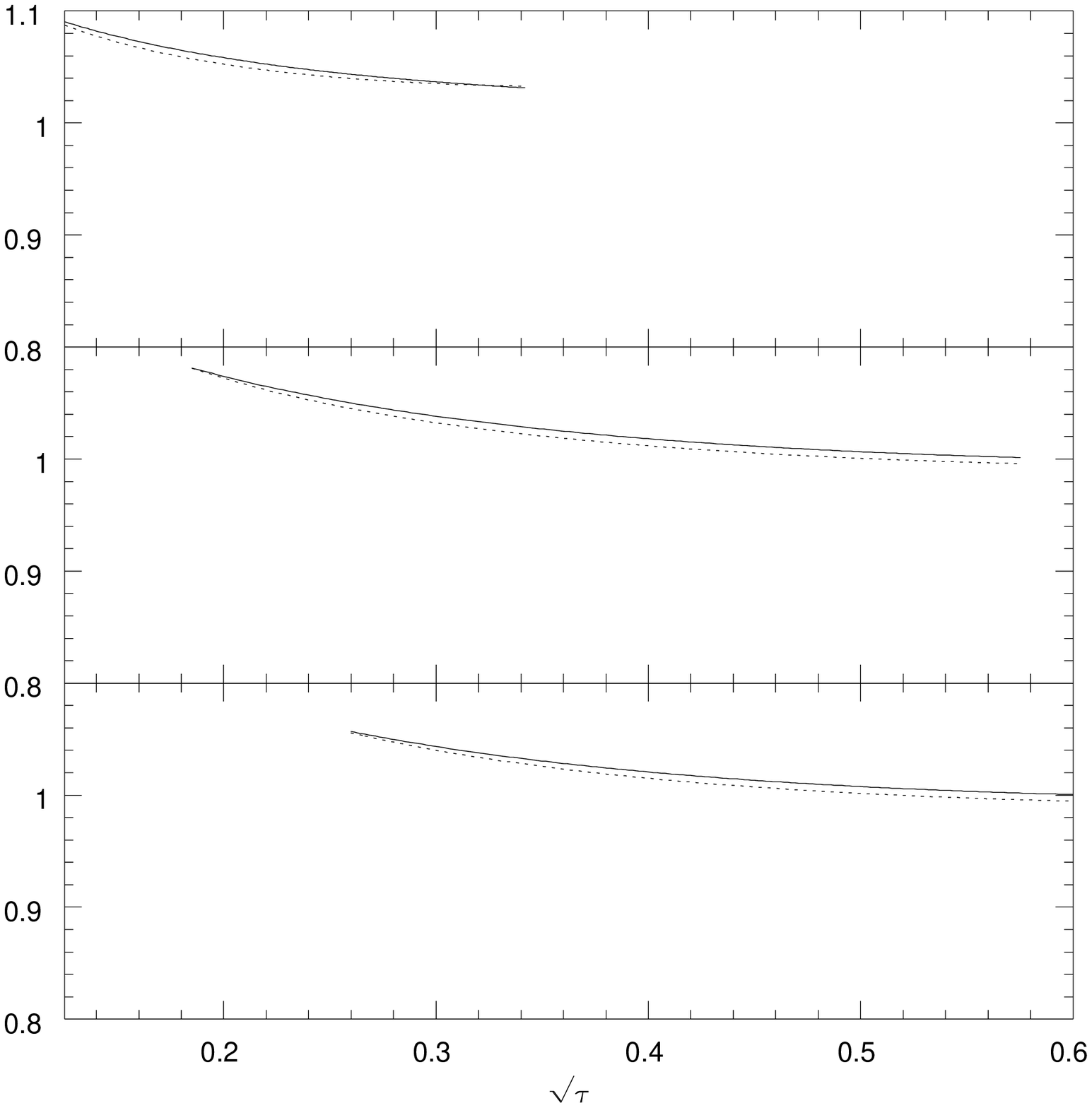}
  \caption{}
  \label{all_dis}
\end{figure}

\begin{figure}
  \epsfysize=450pt
  \epsffile{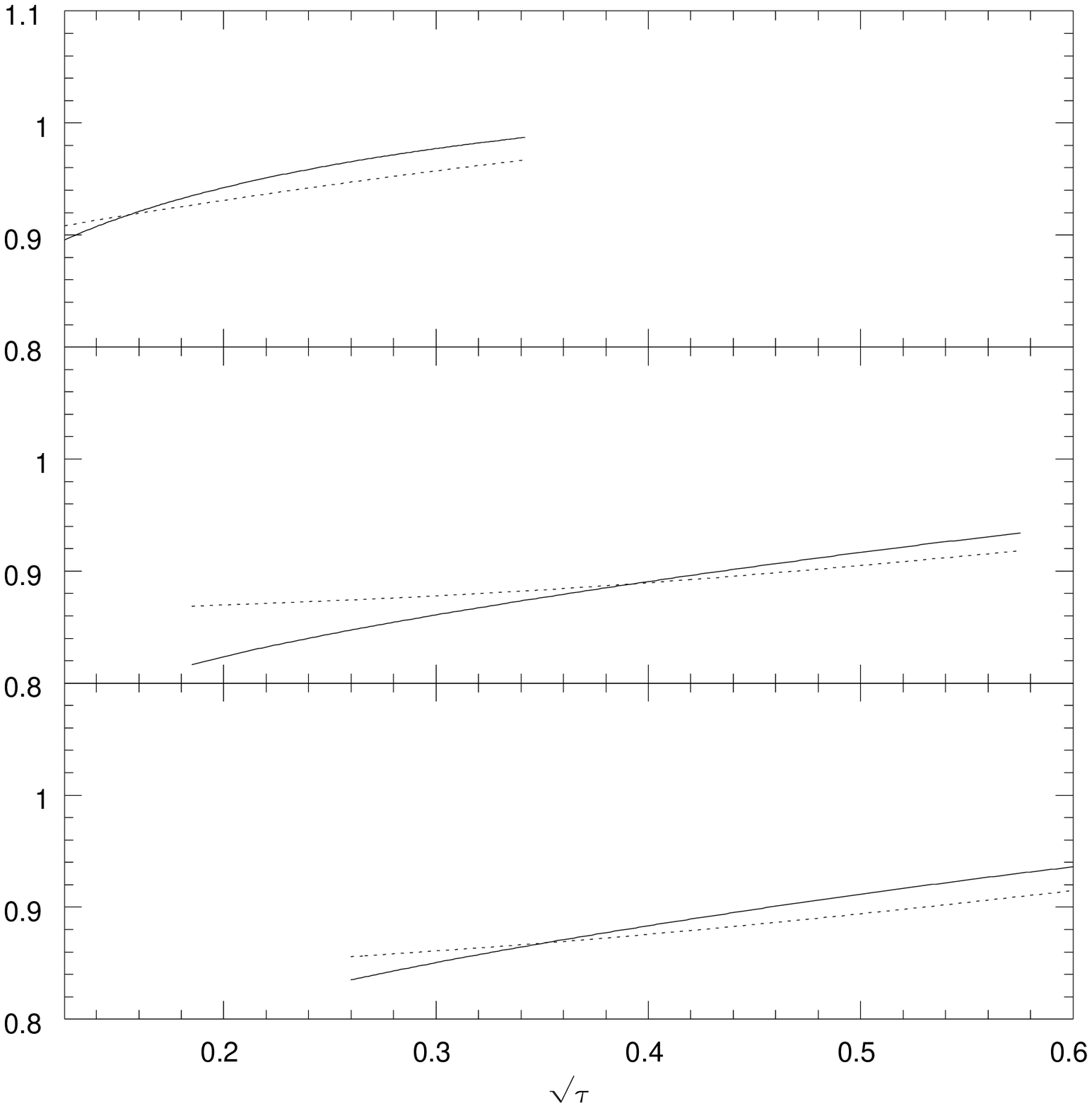}
  \caption{}
  \label{all_ms}
\end{figure}

\begin{figure}
  \epsfysize=450pt
  \epsffile{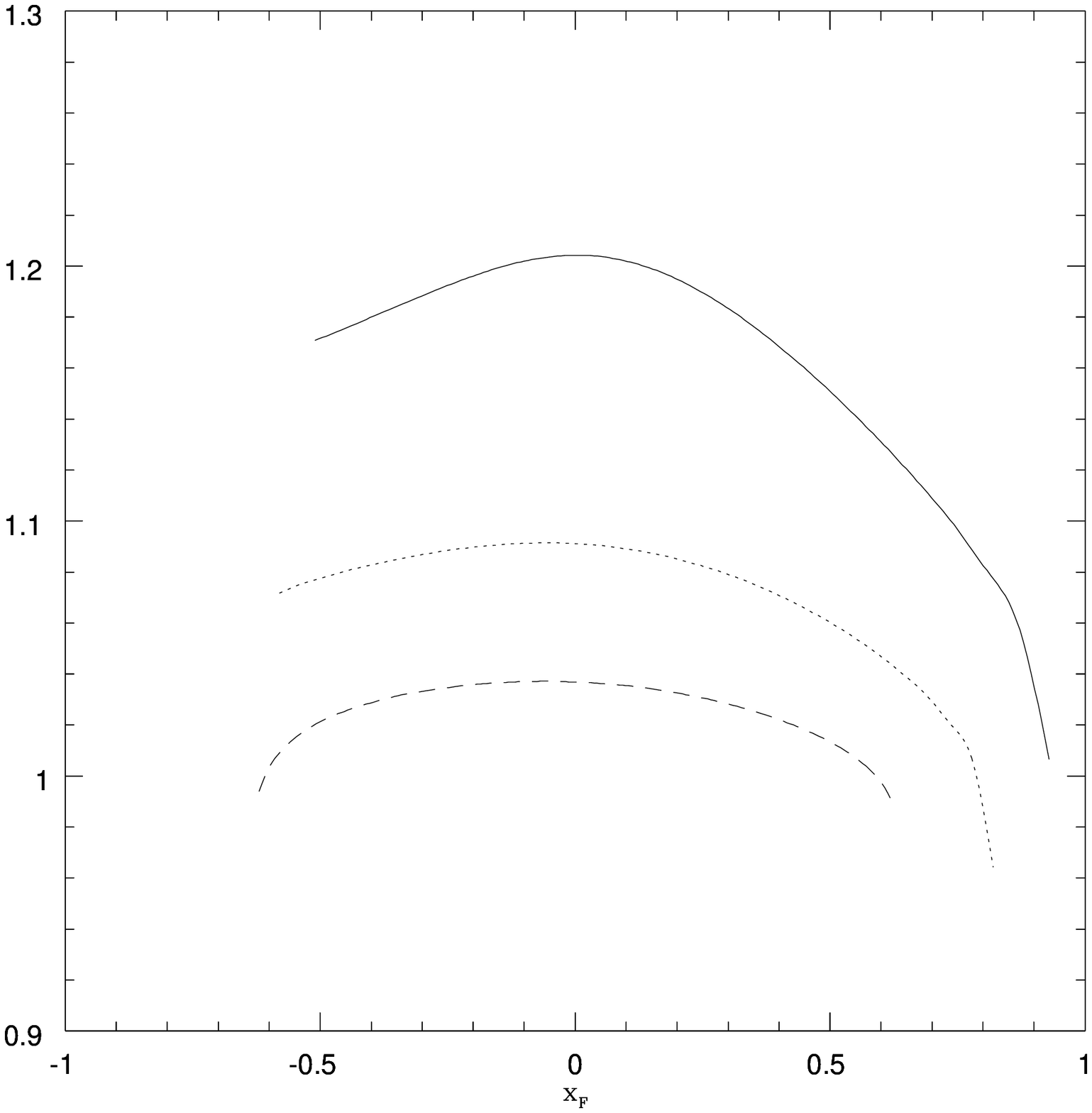}
  \caption{}
  \label{ana_dis}
\end{figure}

\begin{figure}
  \epsfysize=450pt
  \epsffile{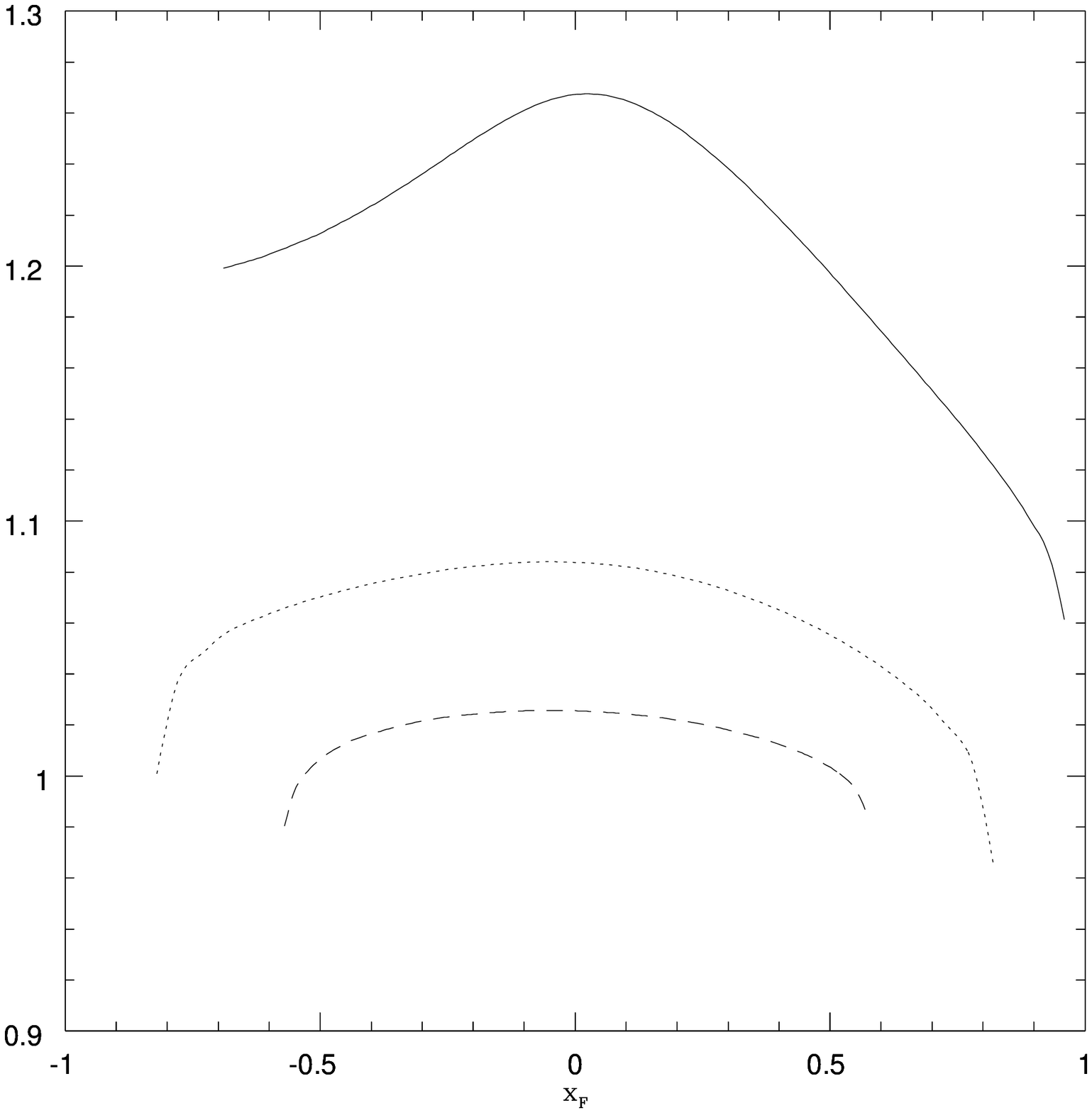}
  \caption{}
  \label{con_dis}
\end{figure}

\begin{figure}
  \epsfysize=450pt
  \epsffile{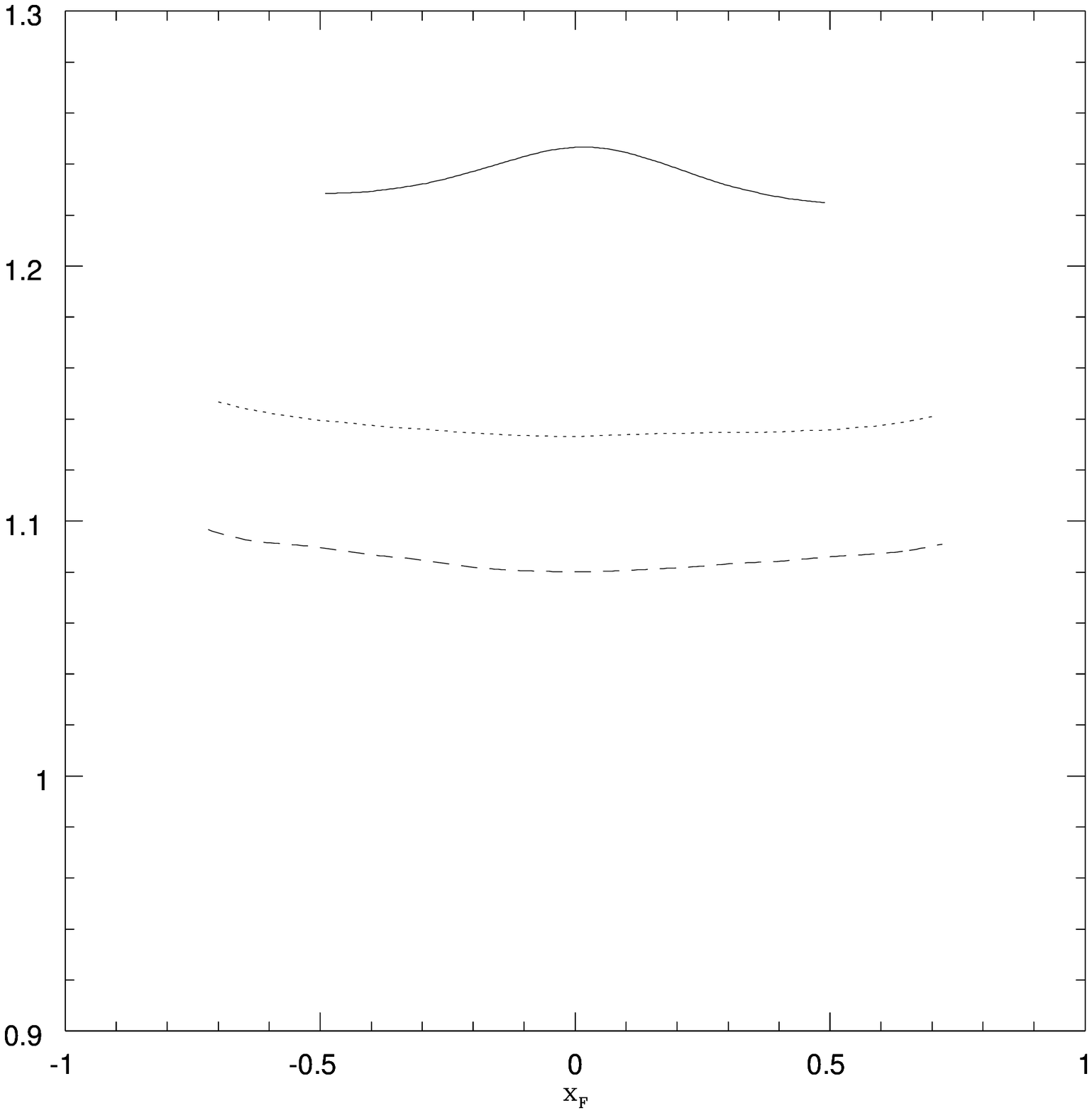}
  \caption{}
  \label{mcg_dis}
\end{figure}

\begin{figure}
  \epsfysize=450pt
  \epsffile{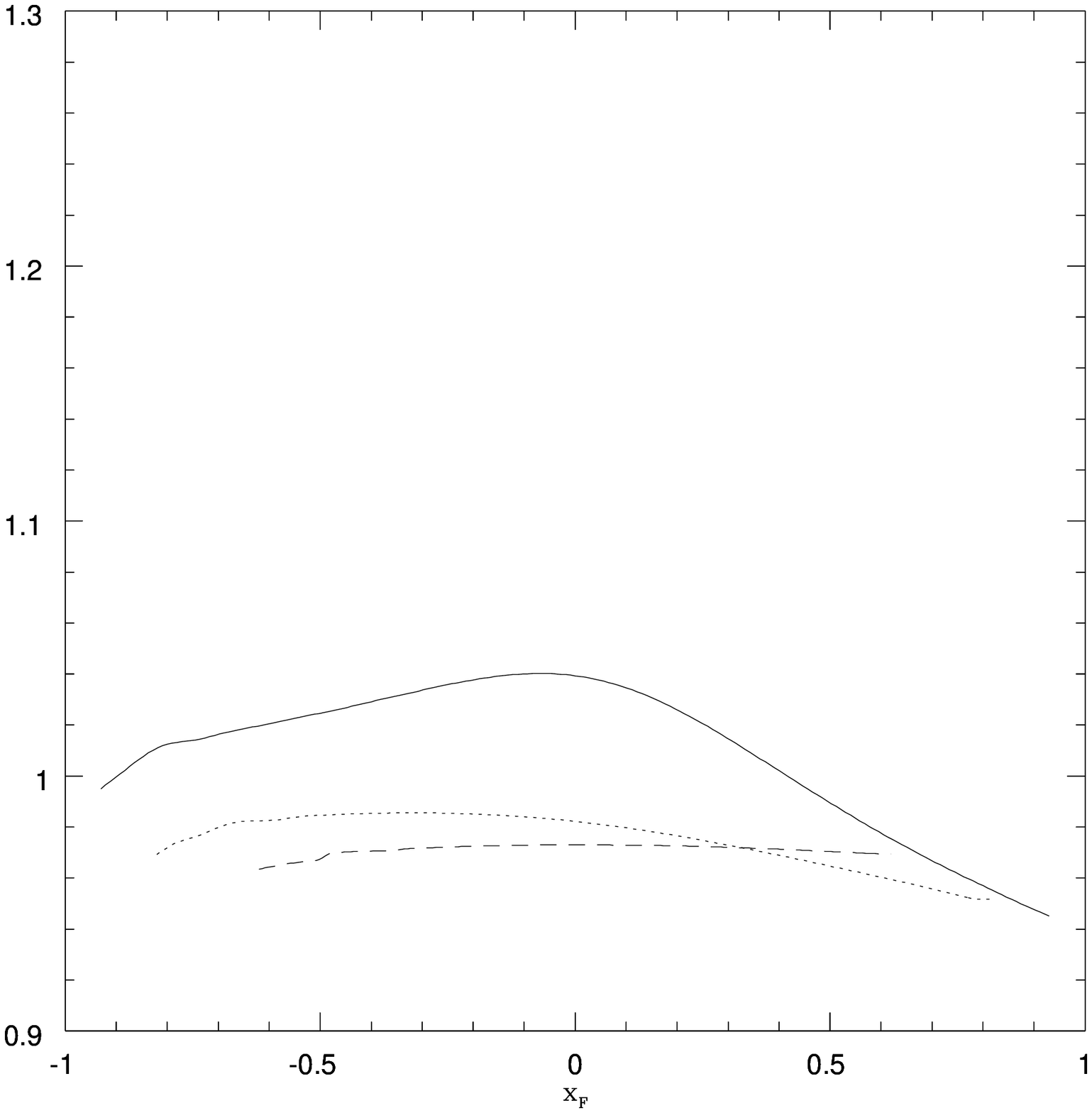}
  \caption{}
  \label{ana_ms}
\end{figure}

\newpage
{\bf Figure captions}
\begin{description}
\item[Fig. 1]
  The ratios $R^{(1)}(\sqrt{\tau})$ (\ref{r1vs}) and $R^{(2)}(\sqrt{\tau})$
  (\ref{r2vs}) presented in the DIS-scheme. Solid line: $R^{(1)}(\sqrt{\tau})$;
  dotted line: $R^{(2)}(\sqrt{\tau})$. Top window : $\sqrt{S} =
  38.8\,\mbox{GeV}/c$ ($0.125 < \sqrt{\tau} < 0.342$, E772). Middle window:
  $\sqrt{S} = 21.8\,\mbox{GeV}/c$ ($0.185 < \sqrt{\tau} < 0.575$, E615). Bottom
  window: $\sqrt{S} = 15.4\,\mbox{GeV}/c$ ($0.26 < \sqrt{\tau} < 0.60$, E537).
\item[Fig. 2]
  The same as in fig.~\ref{all_dis} but now for the
  $\overline{\mbox{MS}}$-scheme.
\item[Fig. 3]
  The ratio $d\,R^{(1)}(\sqrt{\tau},x_{F})$ (\ref{dr1vs}) presented in the
  DIS-scheme for $\pi^{-}+W\rightarrow \mu^{+}\mu^{-} + "X"$ at
  $\sqrt{S}=15.4\,\mbox{GeV}/c$ (E537). Solid line: $\sqrt{\tau}=0.25$; dotted
  line: $\sqrt{\tau} = 0.42$; dashed line: $\sqrt{\tau} = 0.60$.
\item[Fig. 4]
  The ratio $d\,R^{(1)}(\sqrt{\tau},x_{F})$ (\ref{dr1vs}) presented in the
  DIS-scheme for $\pi^{-}+W\rightarrow \mu^{+}\mu^{-} + "X"$ at
  $\sqrt{S}=21.8\,\mbox{GeV}/c$ (E615). Solid line: $\sqrt{\tau}=0.18$; dotted
  line: $\sqrt{\tau} = 0.42$; dashed line: $\sqrt{\tau} = 0.65$.
\item[Fig. 5]
  The ratio $d\,R^{(1)}(\sqrt{\tau},x_{F})$ (\ref{dr1vs}) presented in the
  DIS-scheme for $p+\mbox{}^{2}H\rightarrow \mu^{+}\mu^{-} + "X"$ at
  $\sqrt{S}=38.8\,\mbox{GeV}/c$ (E772). Solid line: $\sqrt{\tau}=0.13$; dotted
  line: $\sqrt{\tau} = 0.23$; dashed line: $\sqrt{\tau} = 0.34$.
\item[Fig. 6]
  The same as in fig.~\ref{ana_dis} but now for the
  $\overline{\mbox{MS}}$-scheme.
\item[Fig. 7]
  The same as in fig.~\ref{con_dis} but now for the
  $\overline{\mbox{MS}}$-scheme.
\item[Fig. 8]
  The same as in fig.~\ref{mcg_dis} but now for the
  $\overline{\mbox{MS}}$-scheme.
\item[Fig. 9]
  $d\sigma/dm$ (\ref{dsdm_half}) for the reaction $\bar{p}+W\rightarrow
  \mu^{+}\mu^{-} + "X"$ at $\sqrt{S}=15.4\,\mbox{GeV}/c$. The data are obtained
  from the E537 experiment \cite{r19}. Dashed line: Born; dotted line:
  $O(\alpha_{s})$ (DIS); solid line: $O(\alpha_{s}^{2})$ (DIS); long dashed
  line: $O(\alpha_{s}^{2})$ ($\overline{\mbox{MS}}$).
\item[Fig. 10]
  The same as in fig.~\ref{dsdm_pbW} but now for the reaction $\pi^{-}+W
  \rightarrow\mu^{+}\mu^{-} + "X"$.
\item[Fig. 11]
  $d\sigma/dx_{F}$ (\ref{dsdxf}) for the reaction $\bar{p}+W\rightarrow
  \mu^{+}\mu^{-} + "X"$ at $\sqrt{S}=15.4\,\mbox{GeV}/c$. The data are obtained
  from the E537 experiment \cite{r19}. Dashed line: Born; dotted line:
  $O(\alpha_{s})$ (DIS); solid line: $O(\alpha_{s}^{2})$ (DIS); long dashed
  line: $O(\alpha_{s}^{2})$ ($\overline{\mbox{MS}}$).
\item[Fig. 12]
  The same as in fig.~\ref{dsdxf_pbW} but now for the reaction $\pi^{-}+W
  \rightarrow\mu^{+}\mu^{-} + "X"$.
\item[Fig. 13]
  $d\sigma/d\sqrt{\tau} = \sqrt{S}d\sigma/dm$ (see (\ref{dsdm_half})) for the
  reaction $\pi^{-}+W\rightarrow\mu^{+}\mu^{-} + "X"$ at $\sqrt{S} = 21.8\,
  \mbox{GeV}/c$. The data are obtained from the E615 experiment \cite{r20}.
  Dashed line: Born; dotted line: $O(\alpha_{s})$ (DIS); solid line:
  $O(\alpha_{s}^{2})$ (DIS); long dashed line: $O(\alpha_{s}^{2})$
  ($\overline{\mbox{MS}}$).
\item[Fig. 14]
  $d^{2}\bar{\sigma}/d\sqrt{\tau}dx_{F}$ (\ref{cross_avg}) with $0.185 <
  \sqrt{\tau} < 0.231$ for the reaction $\pi^{-}+W\rightarrow\mu^{+}\mu^{-} +
  "X"$ at $\sqrt{S}= 21.8\,\mbox{GeV}/c$. The data are obtained from the E615
  experiment \cite{r20}. Dashed line: Born; dotted line: $O(\alpha_{s})$ (DIS);
  solid line: $O(\alpha_{s}^{2})$ (DIS); long dashed line: $O(\alpha_{s}^{2})$
  ($\overline{\mbox{MS}}$).
\item[Fig. 15]
  The same as in fig.~\ref{con_bin1} but now for $0.231 < \sqrt{\tau} <
  0.277$.
\item[Fig. 16]
  The same as in fig.~\ref{con_bin1} but now for $0.277 < \sqrt{\tau} <
  0.323$.
\item[Fig. 17]
  The same as in fig.~\ref{con_bin1} but now for $0.323 < \sqrt{\tau} <
  0.369$.
\item[Fig. 18]
  The same as in fig.~\ref{con_bin1} but now for $0.369 < \sqrt{\tau} <
  0.484$.
\item[Fig. 19]
  The same as in fig.~\ref{con_bin1} but now for $0.484 < \sqrt{\tau} <
  0.599$.
\item[Fig. 20]
  Order $\alpha_{s}^{i}$ corrected $K$-factor denoted by $K_{i}$ (\ref{Kth})
  compared with the experimental $K$-factor (\ref{Kexp}) for the reaction
  $\pi^{-}+W\rightarrow\mu^{+}\mu^{-}+"X"$ at $\sqrt{S}=21.8\,\mbox{GeV}/c$.
  The data are obtained from the E615 experiment \cite{r20}. Dotted line:
  $K_{1}$ (DIS); solid line: $K_{2}$ (DIS); long dashed line: $K_{2}$
  ($\overline{\mbox{MS}}$).
\item[Fig. 21]
  $m^{3}\,d^{2}\sigma/dmdx_{F}$ for the reaction $p+\mbox{}^{2}H\rightarrow
  \mu^{+}\mu^{-}+"X"$ at $\sqrt{S}=38.8\,\mbox{GeV}/c$ and $m=8.15\,
  \mbox{GeV}/c^2$. The data are obtained from the E772 experiment \cite{r22}.
  Dashed line: Born; dotted line: $O(\alpha_{s})$ (DIS); solid line:
  $O(\alpha_{s}^{2})$ (DIS); long dashed line: $O(\alpha_{s}^{2})$
  ($\overline{\mbox{MS}}$).
\item[Fig. 22]
  Parton density dependence of
  $m^{3}\,d^{2}\sigma/dmdx_{F}$ corrected up to order $\alpha_{s}^{2}$ for the
  reaction $p+\mbox{}^{2}H\rightarrow\mu^{+}\mu^{-}+"X"$ at $\sqrt{S}=38.8
  \,\mbox{GeV}/c$ and $m=8.15\,\mbox{GeV}/c^2$.
  The data are obtained from the E772
  experiment \cite{r22}. Solid line: MRS(S0); dotted line: MRS(D0); dashed
  line: MRS(D-).
\end{description}
\newpage

\begin{figure}
  \epsfysize=450pt
  \epsffile{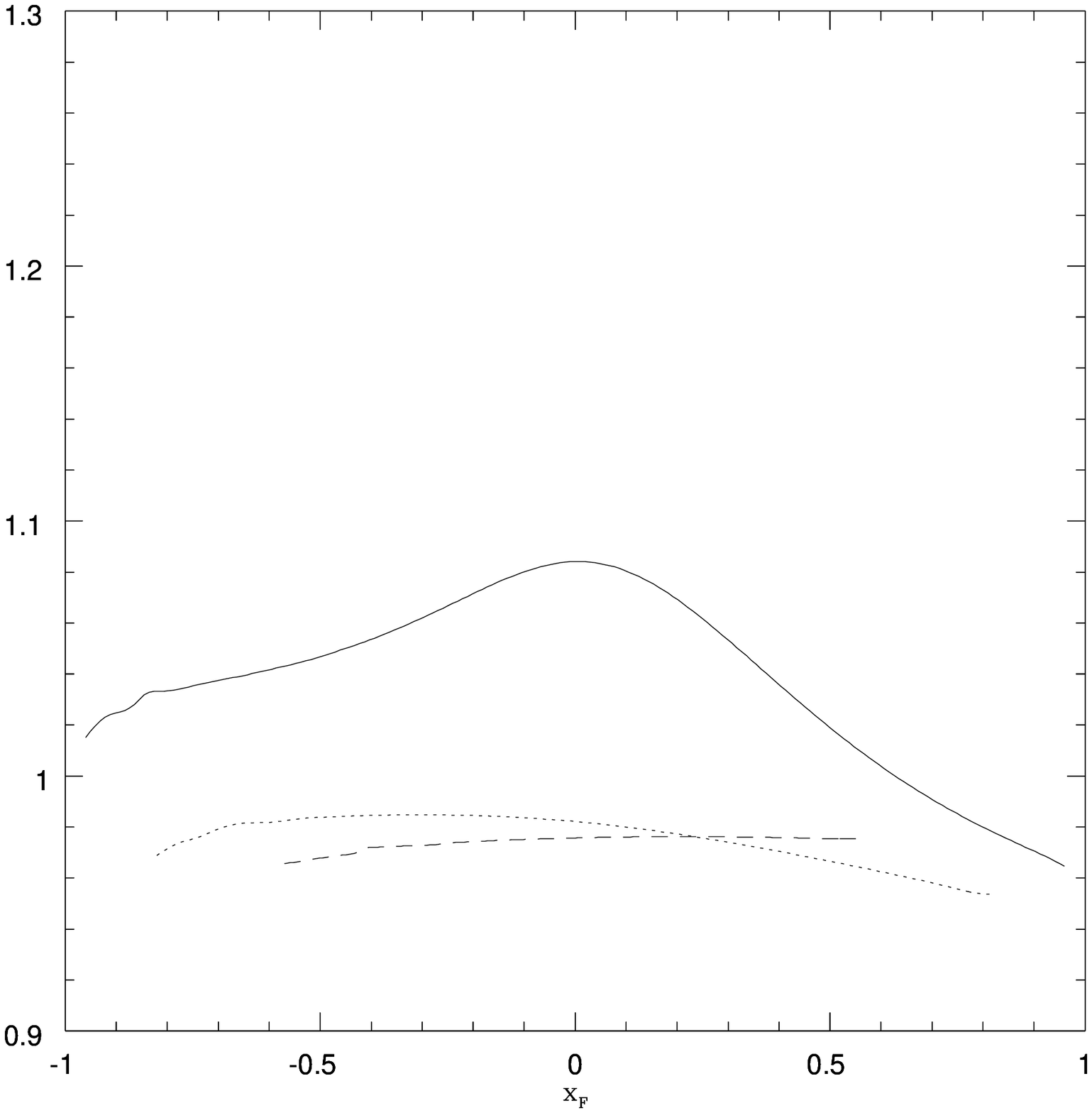}
  \caption{}
\end{figure}

\begin{figure}
  \epsfysize=450pt
  \epsffile{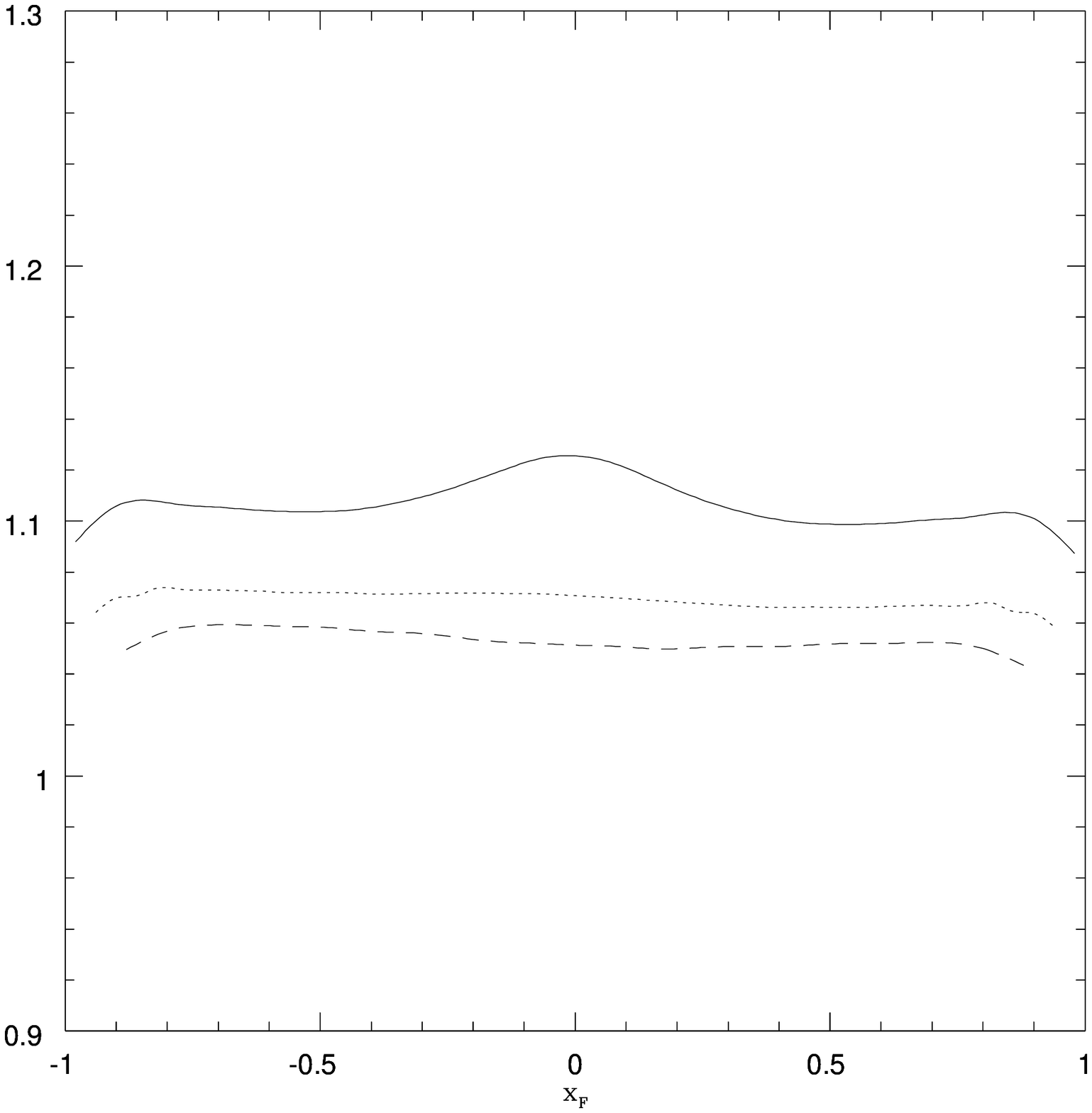}
  \caption{}
  \label{mcg_ms}
\end{figure}

\begin{figure}
  \epsfysize=450pt
  \epsffile{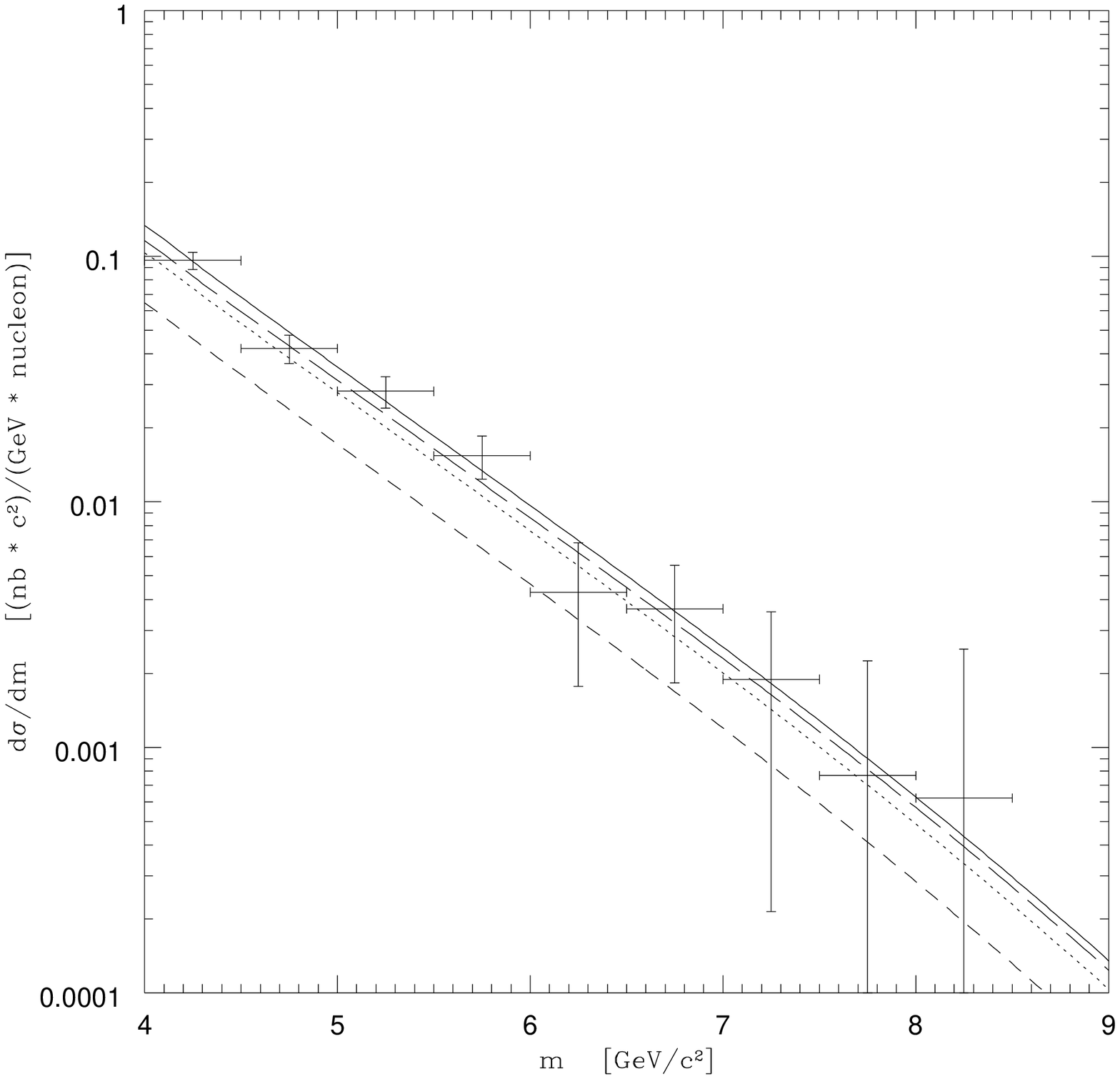}
  \caption{}
  \label{dsdm_pbW}
\end{figure}

\begin{figure}
  \epsfysize=450pt
  \epsffile{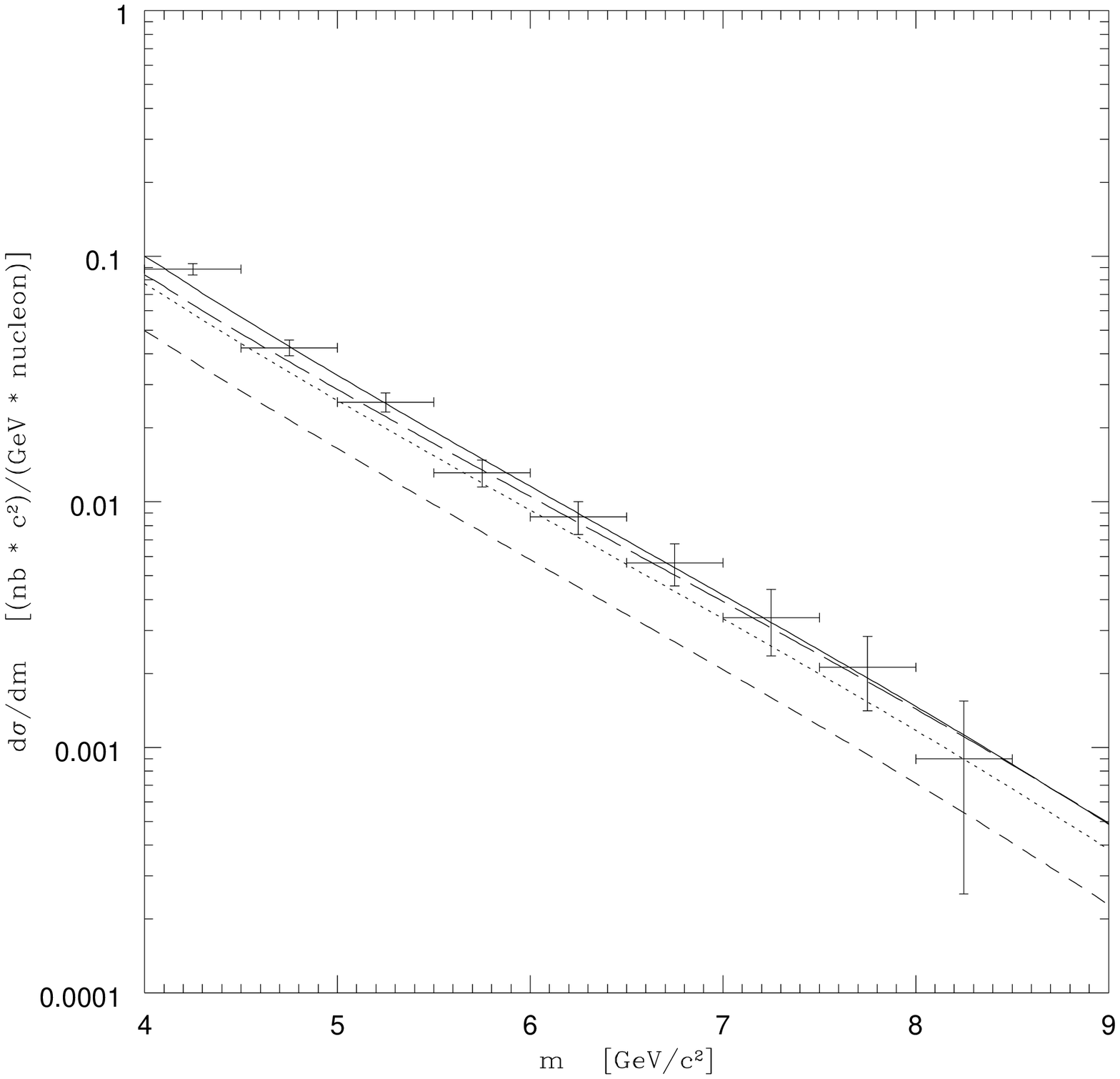}
  \caption{}
  \label{dsdm_piW}
\end{figure}

\begin{figure}
  \epsfysize=450pt
  \epsffile{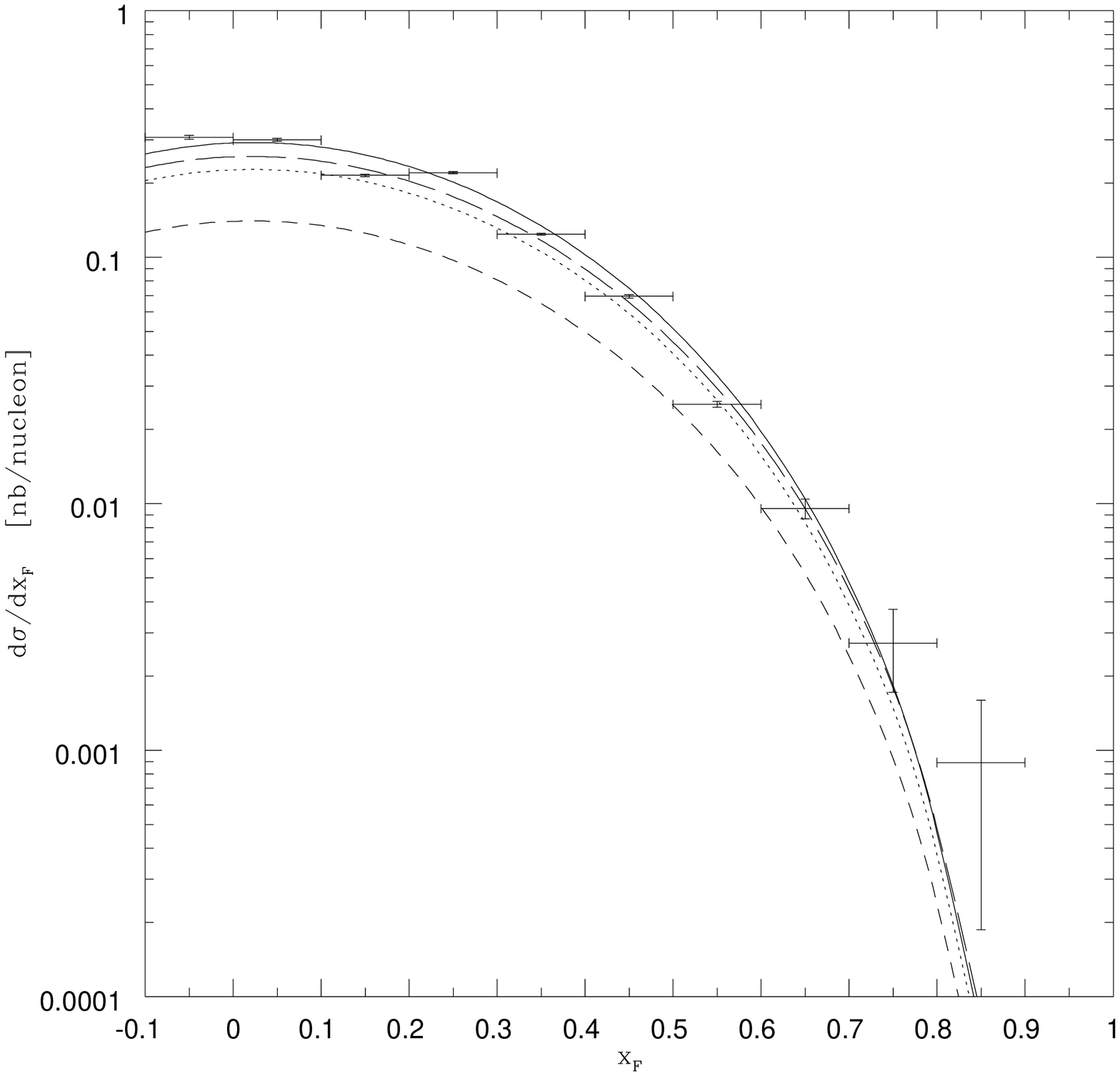}
  \caption{}
  \label{dsdxf_pbW}
\end{figure}

\begin{figure}
  \epsfysize=450pt
  \epsffile{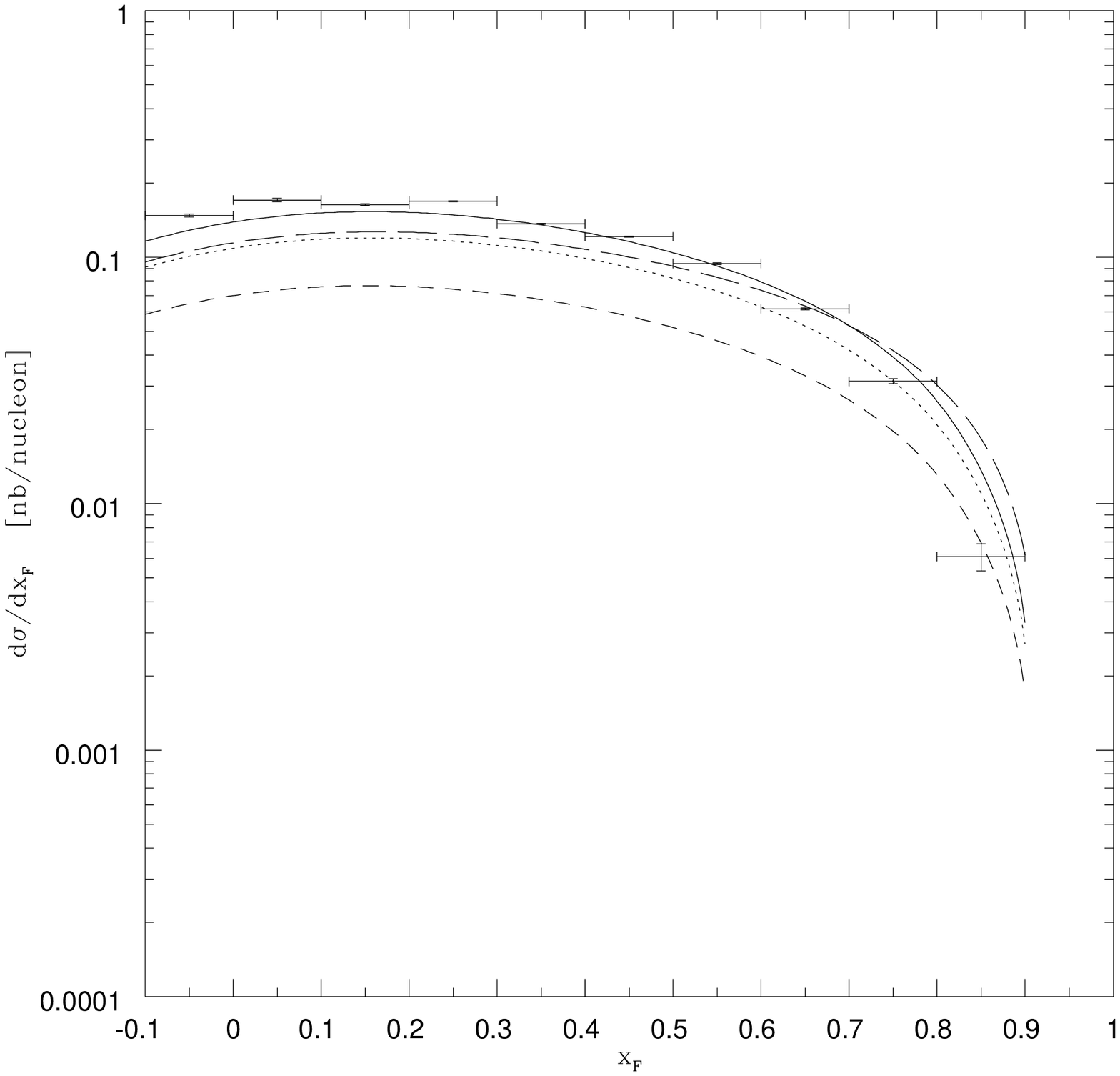}
  \caption{}
  \label{dsdxf_piW}
\end{figure}

\begin{figure}
  \epsfysize=450pt
  \epsffile{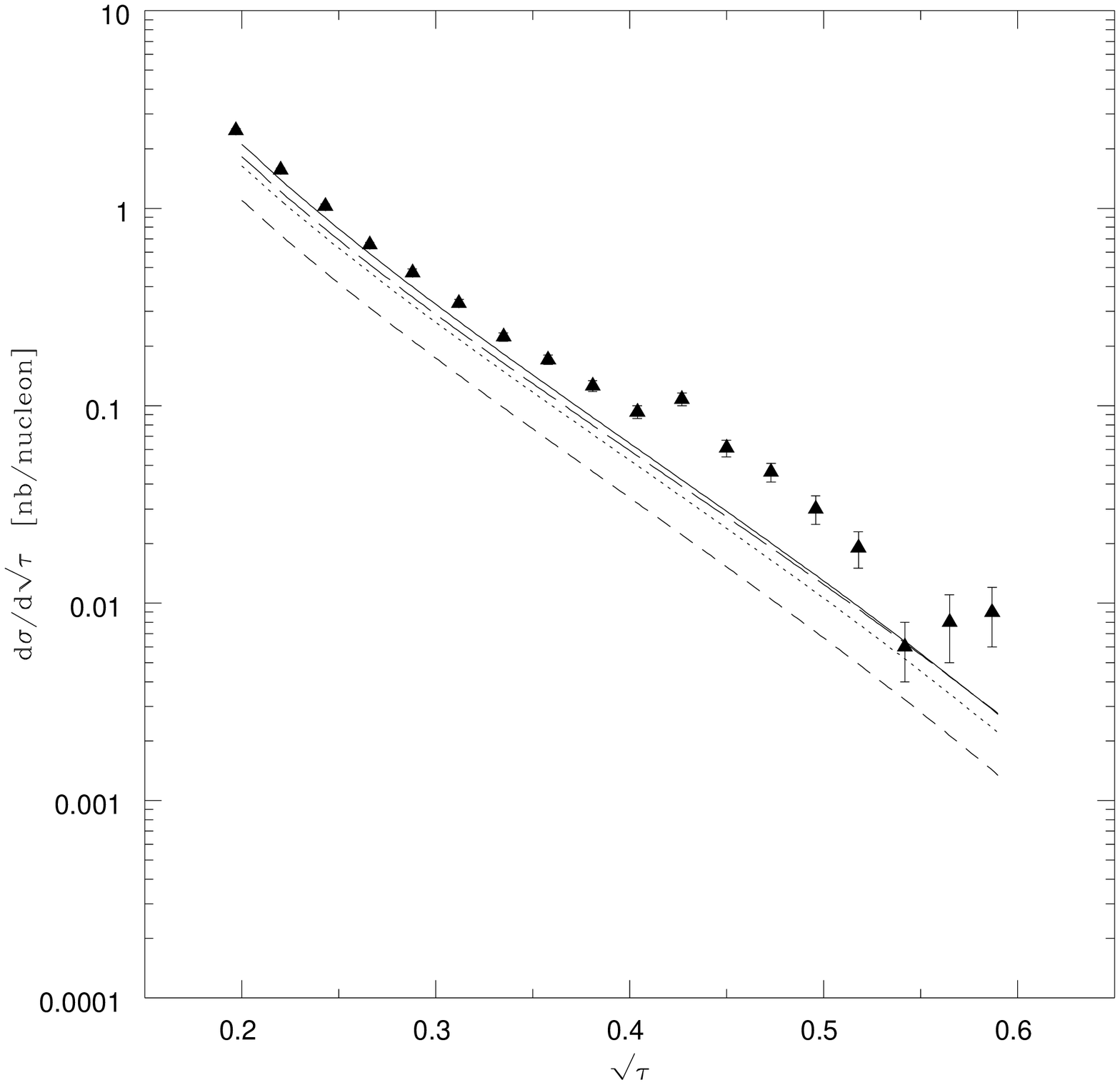}
  \caption{}
  \label{con_dsdm}
\end{figure}

\begin{figure}
  \epsfysize=450pt
  \epsffile{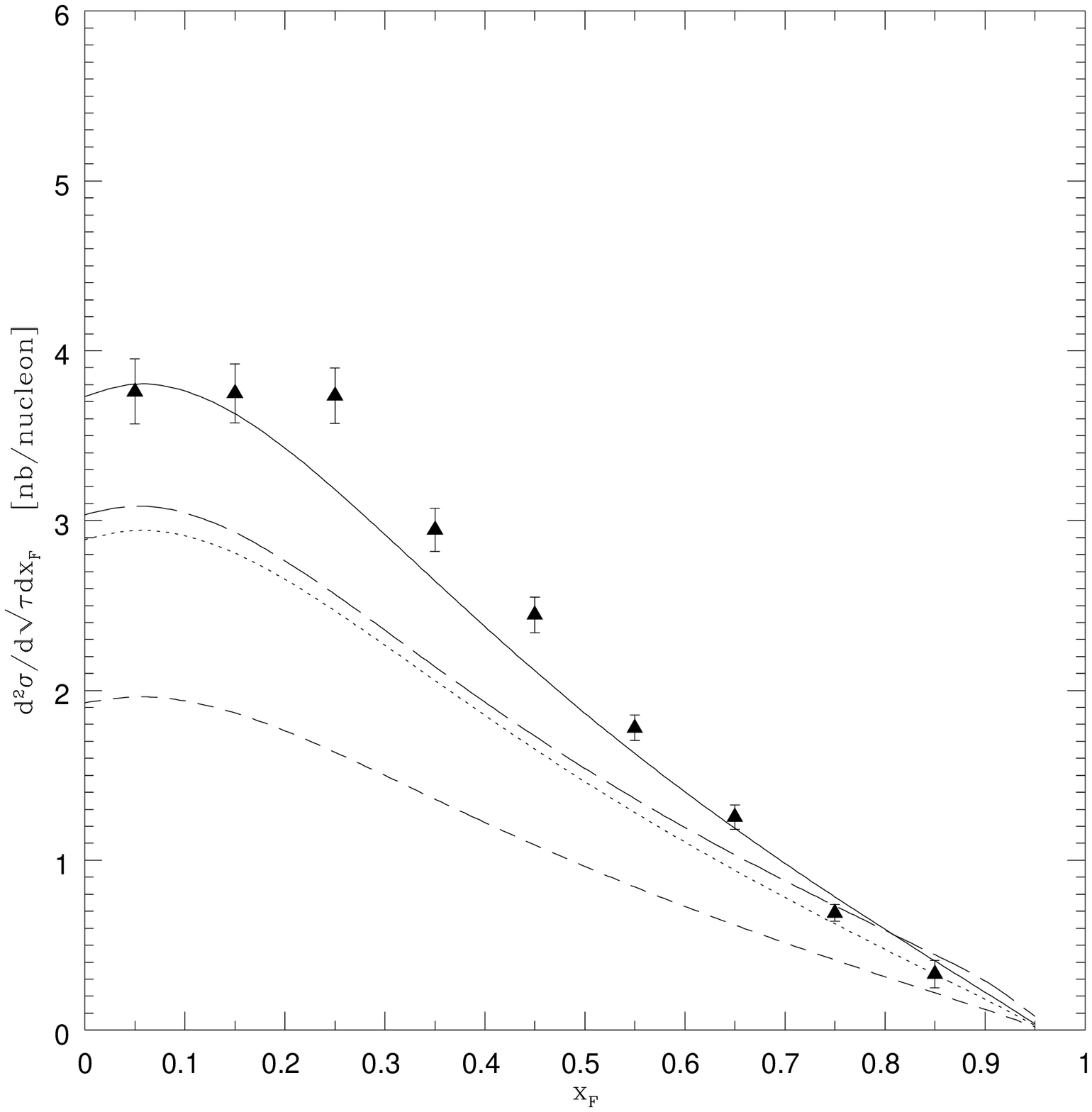}
  \caption{}
  \label{con_bin1}
\end{figure}

\begin{figure}
  \epsfysize=450pt
  \epsffile{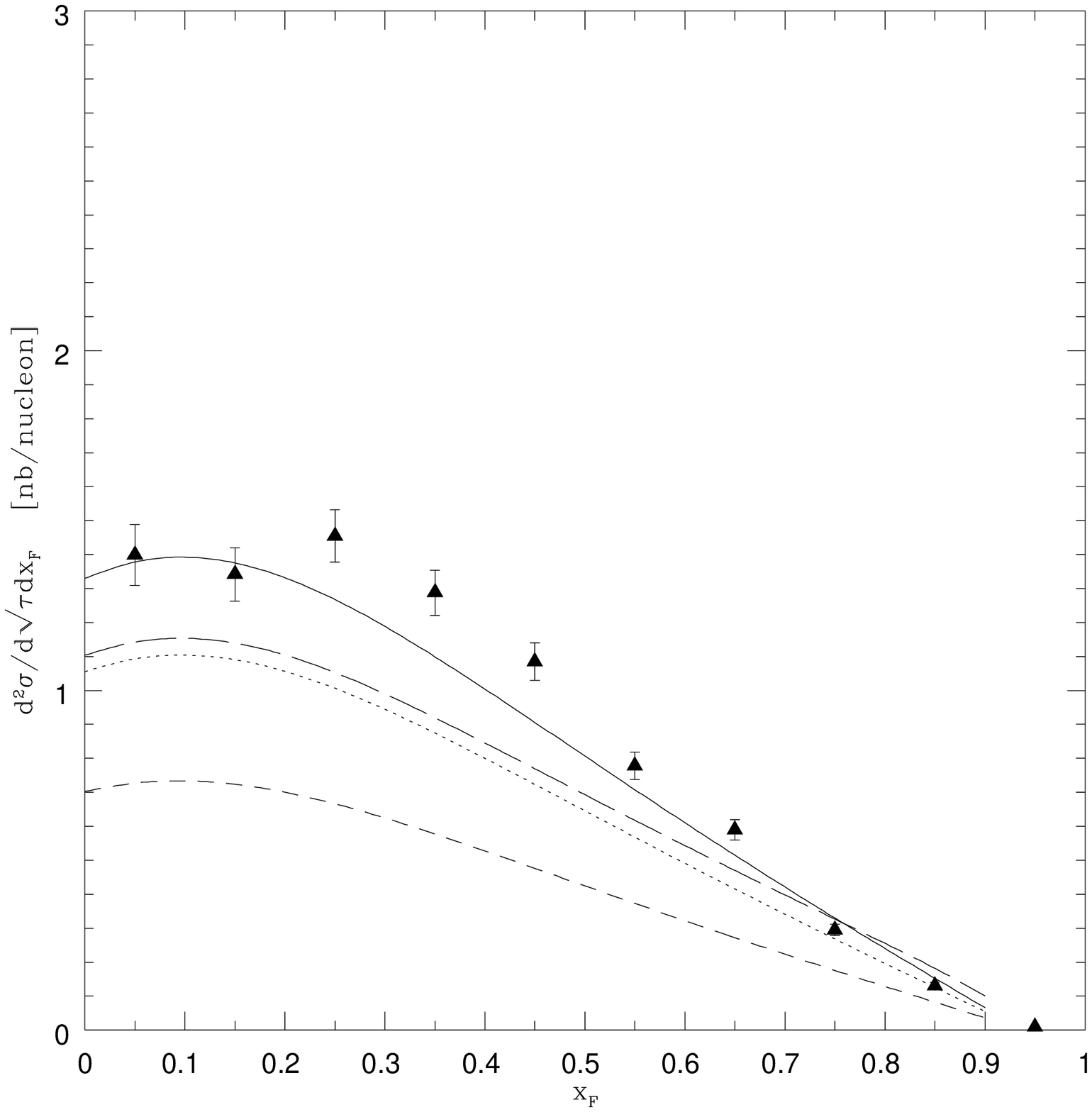}
  \caption{}
\end{figure}

\begin{figure}
  \epsfysize=450pt
  \epsffile{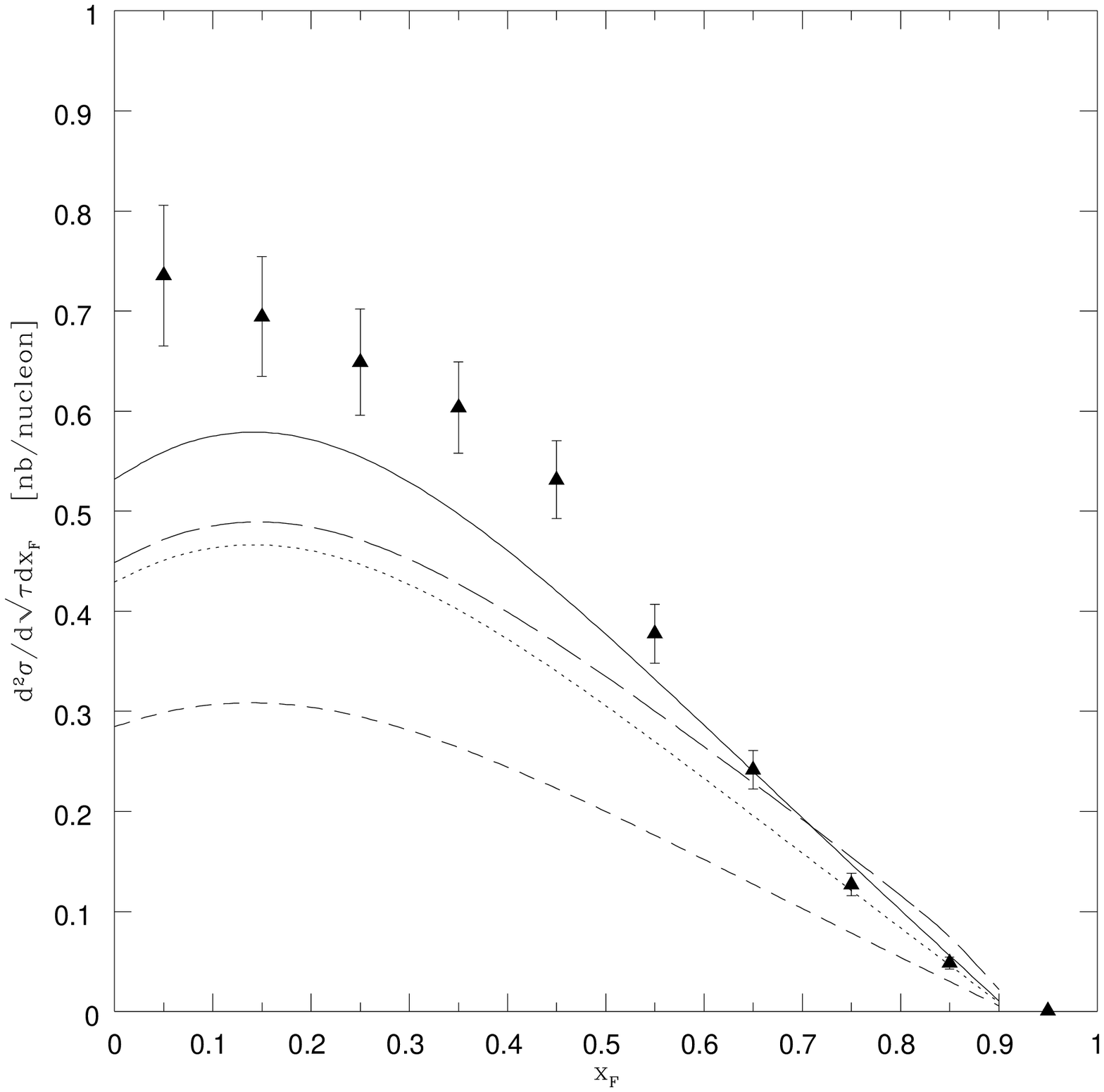}
  \caption{}
  \label{con_bin3}
\end{figure}

\clearpage
\begin{figure}
  \epsfysize=450pt
  \epsffile{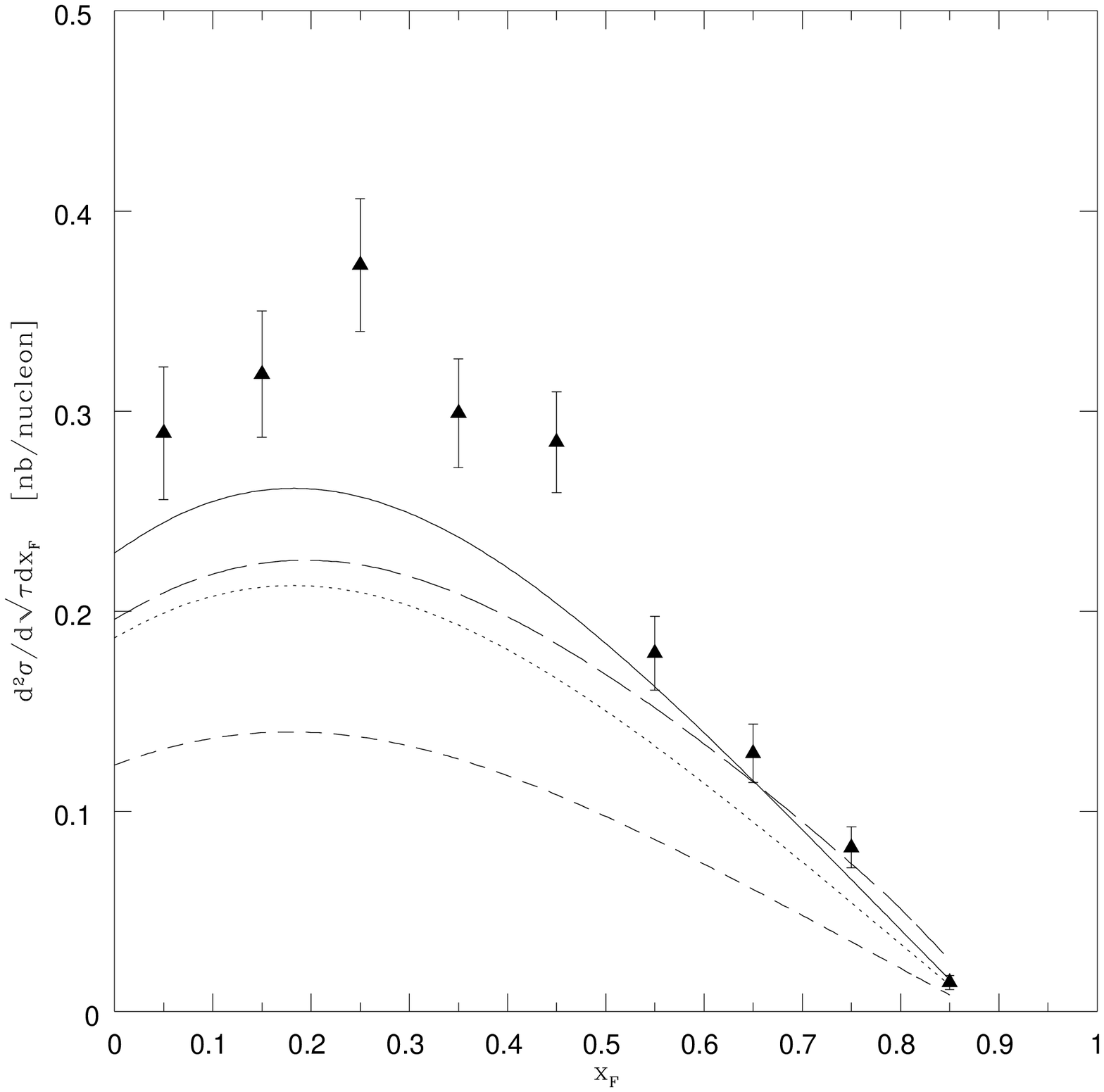}
  \caption{}
  \label{con_bin4}
\end{figure}

\begin{figure}
  \epsfysize=450pt
  \epsffile{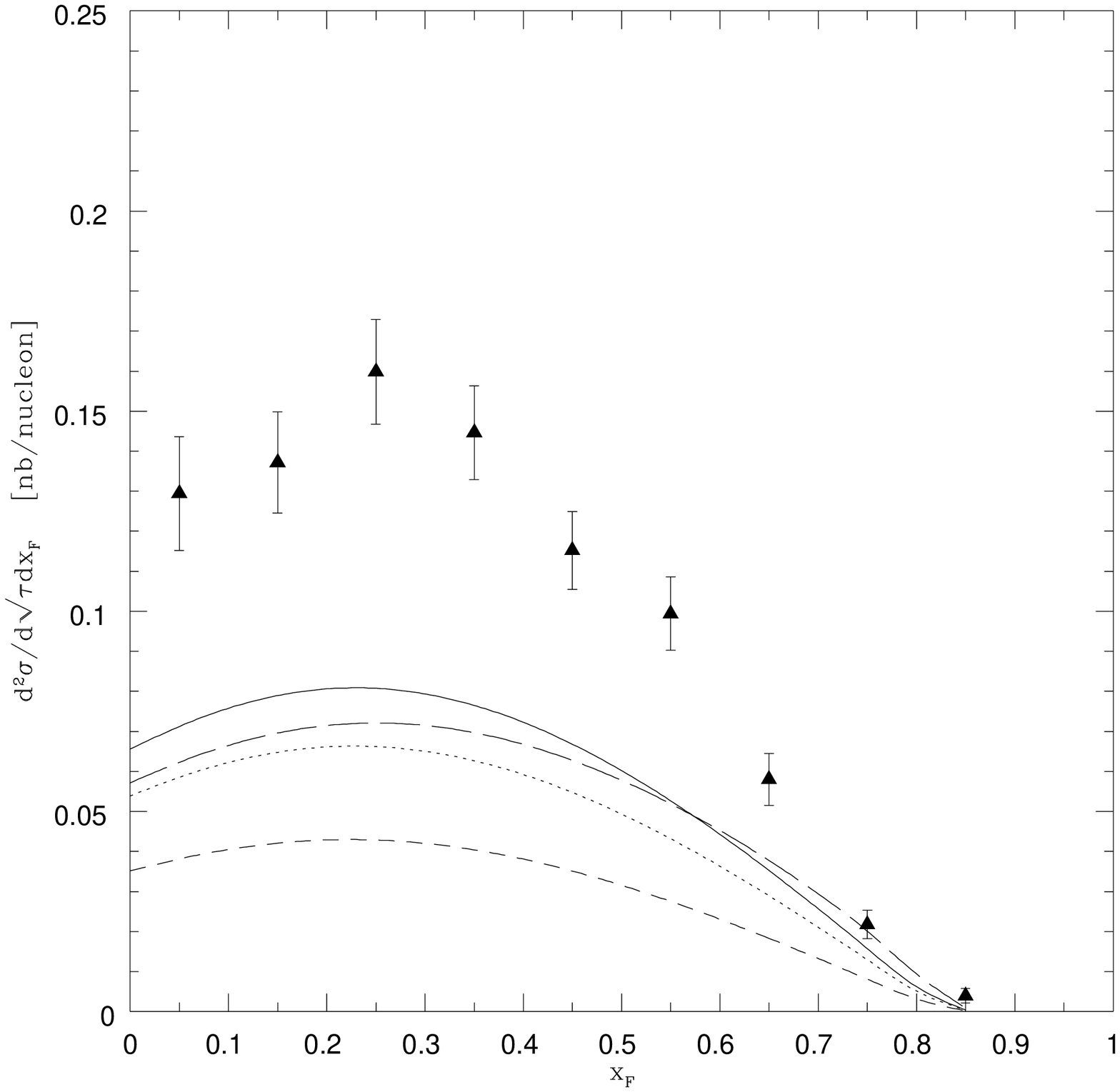}
  \caption{}
\end{figure}

\begin{figure}
  \epsfysize=450pt
  \epsffile{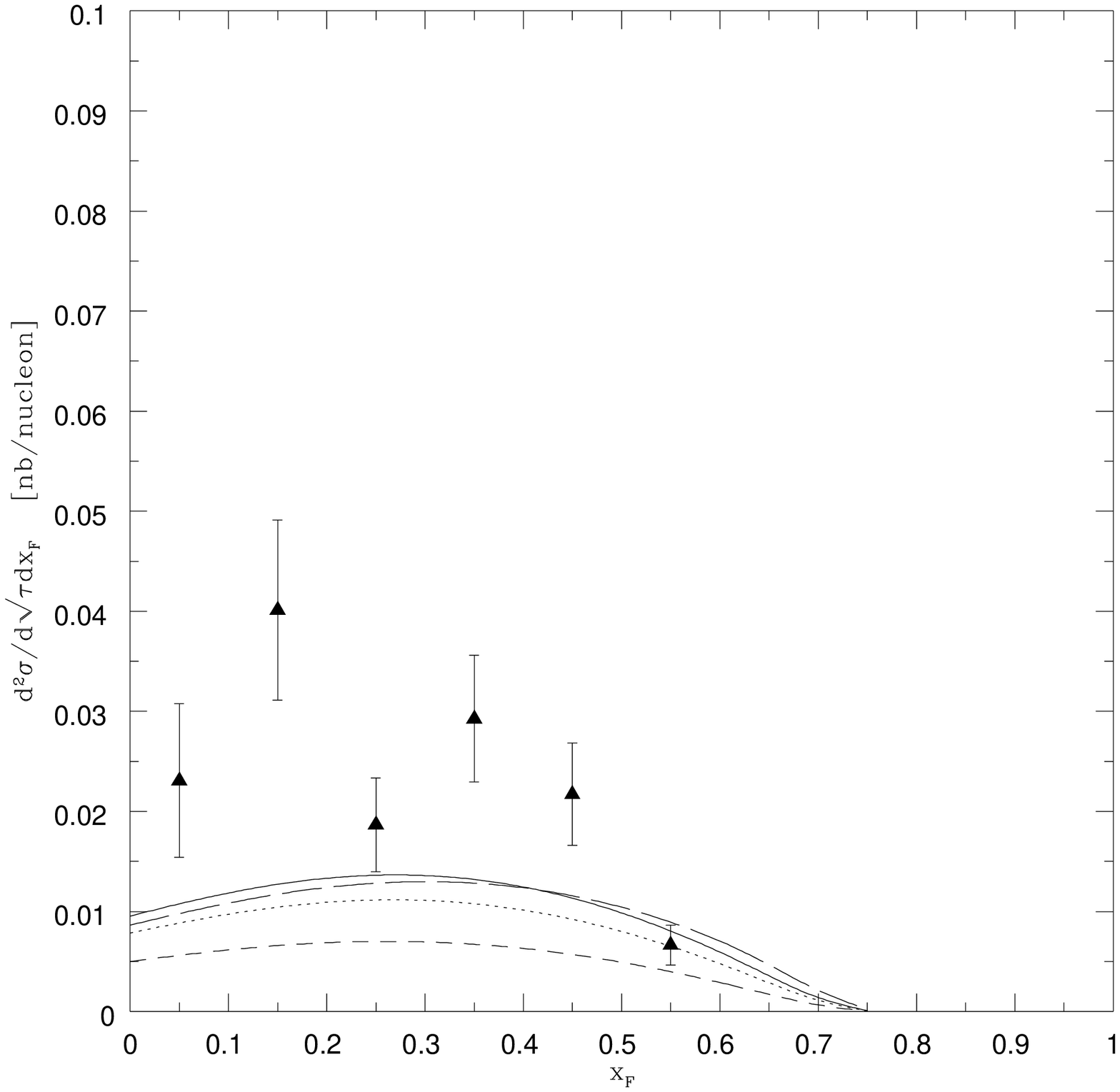}
  \caption{}
  \label{con_bin5}
\end{figure}

\begin{figure}
  \epsfysize=450pt
  \epsffile{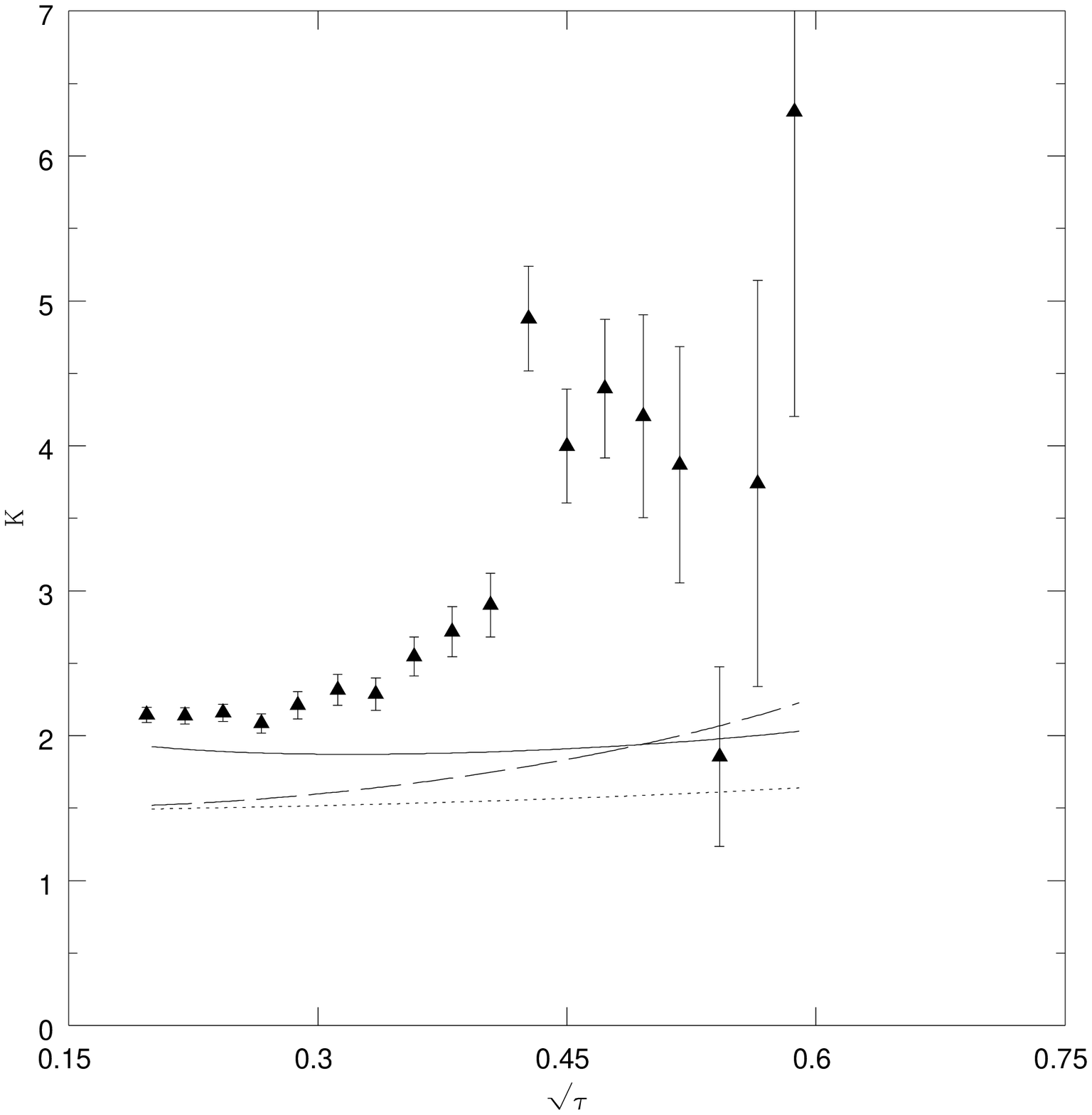}
  \caption{}
  \label{Kfactor}
\end{figure}

\begin{figure}
  \epsfysize=450pt
  \epsffile{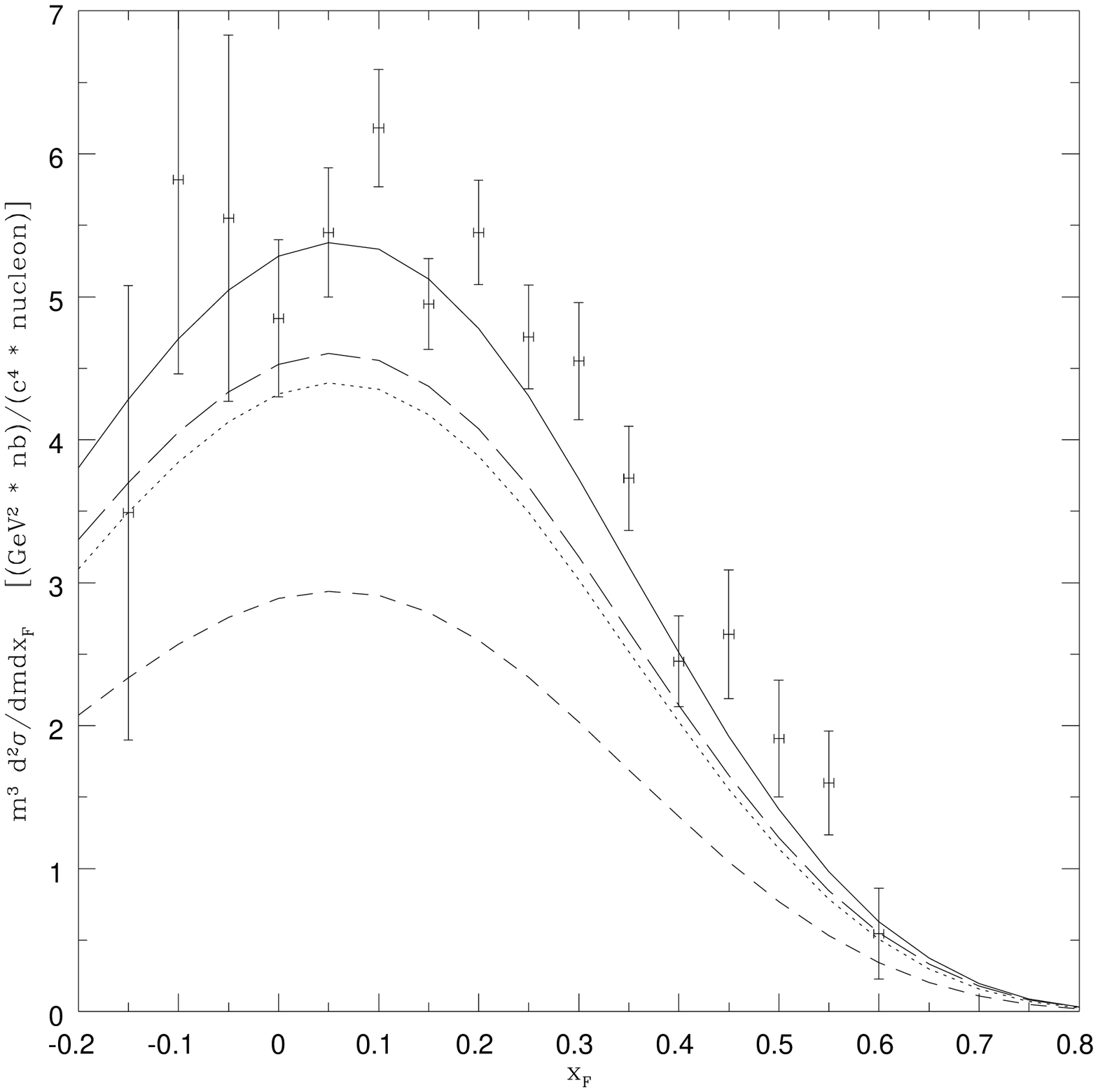}
  \caption{}
  \label{deut_dmin}
\end{figure}

\begin{figure}
  \epsfysize=450pt
  \epsffile{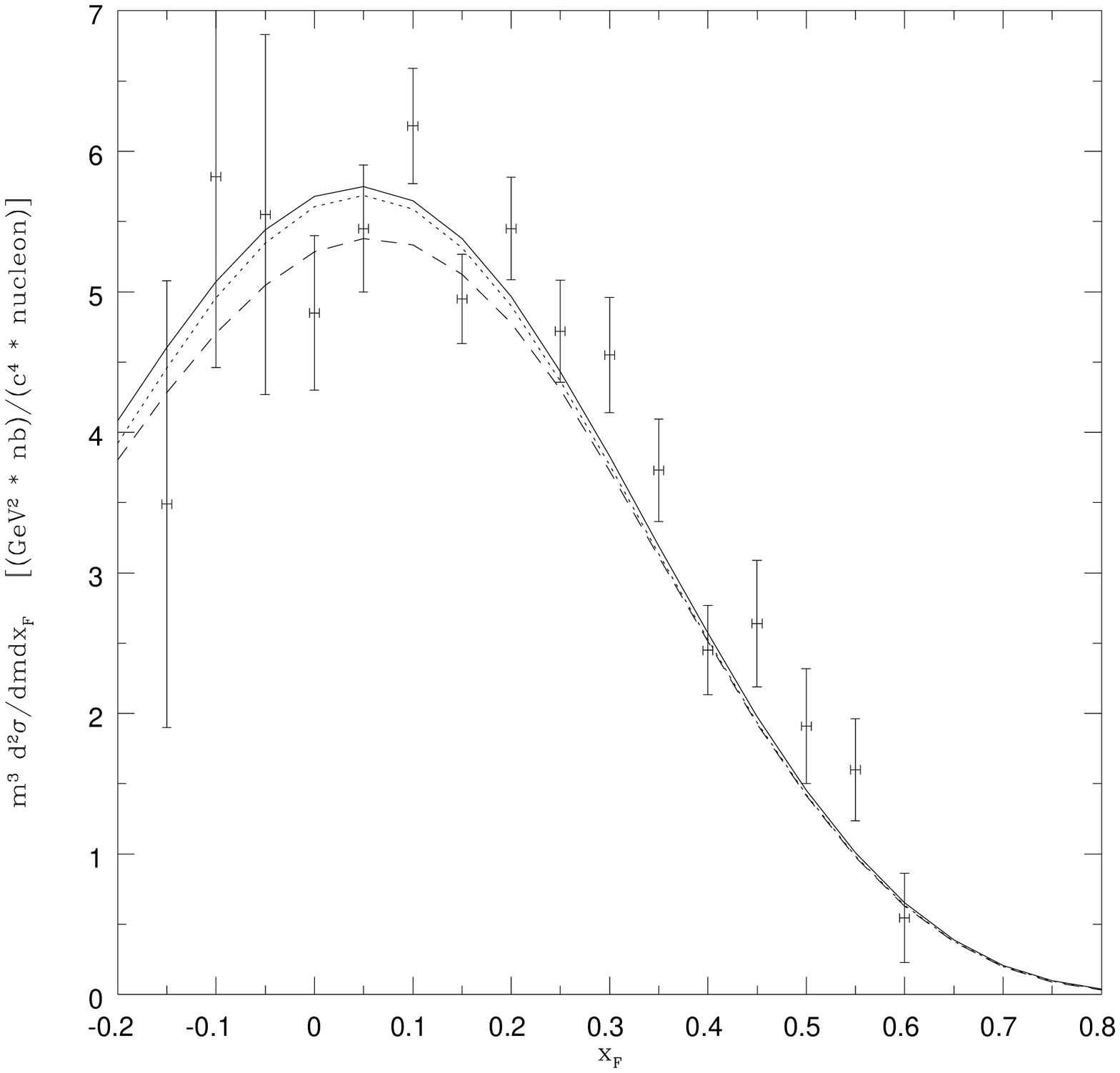}
  \caption{}
  \label{deut_asymm}
\end{figure}
\end{document}